\documentclass[12pt,oneside,a4paper]{article}
\usepackage{epsfig, amsmath,verbatim,cite,graphicx,subeqnarray}
\newcommand{\dis}{\displaystyle}
\newcommand{\expon}{{\rm e}}

\newcommand{\be}{\begin{equation}}
\newcommand{\ee}{\end{equation}}
\newcommand{\bsq}{\begin{subeqnarray}}
\newcommand{\esq}{\end{subeqnarray}}
\newcommand{\bea}{\begin{eqnarray}}
\newcommand{\eea}{\end{eqnarray}}
\begin{document}

%%%%%%%%%%%%%%%%%%%%%%%%%%%%
\renewcommand{\v}[1]{\ensuremath{{\bf #1}}}
\newcommand{\pdif}[2]{\ensuremath{\frac{\partial #1}{\partial #2}}}
\newcommand{\pdiftwo}[2]{\ensuremath{\frac{\partial^2 #1}{\partial
#2^2}}}
\newcommand{\innprod}[2]{\ensuremath{{\langle #1 \mid #2
\rangle}}}
\newcommand{\matel}[3]{\ensuremath{{\langle #1 \mid #2 \mid
#3 \rangle}}} \newcommand{\tendsto}{\ensuremath {\rightarrow}}
\newcommand{\arr}{\longrightarrow}
\renewcommand{\iff}{\ensuremath{\Longleftrightarrow}}
\newcommand{\imp}{\ensuremath{\Longrightarrow}}
\newcommand{\inj}{\ensuremath{\rightarrowtail}}
\newcommand{\surj}{\ensuremath{\twoheadrightarrow}}
\newcommand{\eqrel}{\ensuremath{\!\!\thicksim}}
\newcommand{\xexpr}{\ensuremath{\left( \frac{\delta x}{\epsilon} + r -
\frac{e^y}{2} \right)}}
\newcommand{\yexpr}{\ensuremath{\left( \frac{\delta y}{\epsilon} +
\lambda e^{-y} + \mu - \frac{\xi^2e^{2y(\alpha-1)}}{2} \right)}}
\newcommand{\xexprs}{\ensuremath{\frac{\delta x}{\epsilon} + r -
\frac{e^y}{2}}}
\newcommand{\yexprs}{\ensuremath{\frac{\delta y}{\epsilon} +
\lambda e^{-y} + \mu - \frac{\xi^2e^{2y(\alpha-1)}}{2}}}
\newcommand{\xexpri}{\ensuremath{\left( \frac{\delta x_i}{\epsilon} + r -
\frac{e^{y_i}}{2} \right)}}
\newcommand{\yexpri}{\ensuremath{\left( \frac{\delta y_i}{\epsilon} +
\lambda e^{-{y_i}} + \mu - \frac{\xi^2e^{2y_i(\alpha-1)}}{2}
\right)}}
\newcommand{\xexprsi}{\ensuremath{\frac{\delta x_i}{\epsilon} + r -
\frac{e^{y_i}}{2}}}
\newcommand{\yexprsi}{\ensuremath{\frac{\delta y_i}{\epsilon} +
\lambda e^{-{y_i}} + \mu - \frac{\xi^2e^{2y_i(\alpha-1)}}{2}}}
\newcommand{\bi}[2]{\ensuremath{(#1, #2)}}
\newcommand{\cali}[1]{\ensuremath{\mathcal{#1}}}
\newcommand{\mb}[1]{\ensuremath{\mathbb{#1}}}
%%%%%%%%%%%%%%%%%%%%%%%%%%%%
{\thispagestyle{empty}
\rightline{} %Paper Code Number
\rightline{} %Code Number
\rightline{} %Month, year
\vskip 1cm
%%%%%%%%%%%%%%%%%%%%%%%%  title %%%%%%%%%%%%%%%%%%%%%%%%%%%
\centerline{\large \bf Simulation of Stochastic Volatility }
\centerline{\large \bf using Path Integration: Smiles and Frowns }

%%%%%%%%%%%%%%%%%%%%%%%%%%%%%%%%%%%%%%%%%%%%%%%%%%%%%%%%%%%

\vskip 2cm
%%%%%%%%%%%%% Author  and Address %%%%%%%%%%%%%%%%%%%%%%%%%
\centerline{ Belal E. Baaquie \footnote{Email:phybeb@nus.edu.sg},
L.C. Kwek and M. Srikant} \centerline{ \it Department Of Physics,
National University of Singapore,} \centerline{ \it Kent Ridge,
Singapore 119260 }
%address
\vskip 0.1in
%%%%%%%%%%%%%%%%%%%%%%%%%%%%%%%%%%%%%%%%%%%%%%%%%%%%%%%%%%%

\vskip 1cm
\centerline{\bf Abstract} \vspace{10mm}

\noindent{ We apply path integration techniques to obtain option
pricing with stochastic volatility using a generalized
Black-Scholes equation known as the Merton and Garman equation. We
numerically simulate the option prices using the technique of path
integration. Using market data, we determine the parameters of the
model.  It is found that the market chooses a special class of
models for which a more efficient algorithm, called the bisection
method, is applicable. Using our simulated data, we generate some
implied volatility curves.  We also analyze and study in detail
some of the characteristics of the volatility curves within the
model. } }
\newpage

%%%%%%%%%%%%%%%%%%%%%%% Introduction %%%%%%%%%%%%%%%%%%%%%%%%

\section{Introduction}

Two popular instruments for option valuation in the market are the
binomial option pricing and the Black-Scholes model\cite{black}.
The underlying stochastic nature of  these models resembles
quantum mechanical theories.  In quantum mechanics, tools have
been invented to solve and compute such stochastic quantities; one
widely used instrument in this field is an intriguing integral
called the path integration.  The application of this method to
option pricing has been mooted and elucidated in great detail in
elsewhere \cite{baaq}.  The underlying principle behind the method
is essentially based on a generalized Black-Scholes model but
unlike the Black-Scholes model, the new formalism can easily
accommodate stochastic volatility and provide wider scope and
greater flexibility for the investigation of market behavior.

The most direct advantage of recasting the option pricing problem
in terms of this Feynman path integral is that this formalism
allows a new perspective for the trader and market analysts, and
can lead to several new ways of computing exact, approximate and
numerical solutions for the pricing of option. Stochastic
volatility is naturally incorporated within the model. Stochastic
volatility can introduce a high degree of nonlinearity within the
option pricing problem.  The path integral formalism can in
principle handle such nonlinearity in an elegant manner. Moreover,
there is a possibility that traders who are learnt this new tool
can formulate new exotic options.

There are many inherent similarities between the problem of
pricing options in finance and solution of models in the physical
sciences using stochastic approach.  Indeed, this striking
similarity have prompted Bouchaud and others
\cite{bour,pott1,cont1}  to apply successfully many mathematical
tools previously used by physicists, like functional integration
and scale invariance, to analyze problems in the financial
markets.

The celebrated Black-Scholes equation provides a simple and
analytical formula for traders interested in plain vanilla
European option. Equipped with the formula, analysis of the
simplest option simply involves taking parameters estimated from
historical data and working out the price of the option. Nowadays,
even in the simplest scenario, market analysts utilize option
pricing using Black-Scholes formula with a twist. Instead of
estimating the constant volatility required in Black-Scholes
equation, which is often difficult and inaccurate, they usually
compute the implied volatility which is necessary for the traded
option price for a specific strike price to be consistent with the
Black-Scholes formula. Such an analysis gives rise to a graph of
implied volatility against strike price which normally appears as
a smile or a frown. The decision regarding various investment
portfolios depends largely on the difference between this implied
volatility and the historical volatility. Large implied volatility
is taken by traders to mean that the option is overpriced.

Numerous extensions on the Black-Scholes model have been proposed,
see for instance Wilmott's book on financial derivatives
\cite{wilmott} and references therein.  Moreover, to explain the
deviation of the call option from the Black-Scholes model, many
refinements to the original model have been studied. For instance,
one possible refinement is to relax the fixed volatility rate and
replace it with stochastic volatility. Another refinement is to
consider the stochastic model for the pricing of the security as a
jump-diffusion process \cite{jarrow, cox85}. Recently, it has been
shown by Das and Sundaram\cite{sanjiv} that certain securities
which follow a jump-diffusion process can exhibit stochastic
volatility. There are essential differences between the last two
examples of possible refinements. For the jump-diffusion case, the
effect is enhanced within short terms whereas for stochastic
volatility, the effect is more pronounced over a longer maturity
time.  Finally, one can also combine the two effects and consider
stochastic interest rates.

Constructing models for option pricing with stochastic volatility
is an important issue since there are strong indications of
stochastic variations in the underlying asset pricing and their
derivatives from numerous empirical data. Several papers have
recently attempted to study this pricing bias caused by stochastic
volatility \cite{hull2,stein,ball,heston, sanjiv}. These studies
have generally focused on numerical solutions of partial
differential equations, Fourier inversion methods and the power
series approximation techniques.  The method of solving partial
differential equations to study the option pricing has been
considered to be the one of the general approach. However, a major
distinct disadvantage of this technique is that most of the work
done using this method is heavily computer intensive.  The path
integral formulation, on the other hand, offers an intermediate
alternatives in many instances. It can include many of these
techniques and offer new insights into the pricing of options.
Moreover, the path integrals can yield results in a global
approach involving the properties of the model at all times.

In this paper, we investigate option valuation using the path
integral approach.  We generate the volatility curves and study
their behaviors.  As the application of path integration to option
pricing differs from the current theoretical method used for
evaluating the option prices in the market, we shall briefly
sketch the theoretical basis of our formulation in the next
section. A detailed review and description of the path integral
formalism in option pricing will not be given here. A more
elaborated account of the path integral formalism to option
pricing can be found in the following reference\cite{baaq}.

In section \ref{theory}, we discuss at length the theoretical
framework of our model.   The general algorithm using Monte Carlo
techniques is given in the section \ref{generalalgo}.  We perform
the numerical simulations for our model in section \ref{gensim}
and calibrate market data to the model in the same section.  We
find that the market seems to choose a special set of parameters.
For this set of parameters, there exist more efficient algorithm
called the bisection method. In section \ref{num}, we describe
this algorithms  and provide the necessary pseudocodes for the
bisection method. We also briefly discuss the convergence of the
program.

Finally, in section \ref{disc}, we briefly discuss and study some
characteristics of the implied volatility curves generated. Our
simulation is based on Merton and Garman equation and
consequently, there are several parameters which we can vary
within the model.  Besides the correlation coefficient between the
security prices and the volatility, our model involves two other
arbitrary constants; one related to the variance and the other
related to the mean of the stochastic volatility.  The initial
stock price and the initial volatility can also be varied and
studied. Using the path integral formalism, we generate computer
data on option pricing by varying the various parameters in the
model. Based on the option prices, we compute the implied
volatilities and  analyze the graphs of implied volatility against
strike price.

\section{Theoretical Formulation}\label{theory}
%%%%%% Srikant's thesis chapter5 %%%%%
Black and Scholes \cite{black} laid the foundation for a
quantitative analysis of European options. Since then, several
extensions have been done and some of the original assumptions
have been dropped.  In an important paper, Merton\cite{merton1}
ingeniously  removed the assumption of constant interest rates and
showed that an option can be priced in terms of a bond price. In
the same paper, Merton also showed how the Black-Scholes formula
can be extended to cover situations in which the volatility is a
deterministic function of time. Indeed, based on a critical
analysis of market data, Rubinstein has shown that the assumption
of  constant volatility is generally incorrect\cite{rubenstein}.
Research has also been done assuming different processes for the
evolution of stock prices by Merton\cite{merton1a}, Cox and
Ross\cite{cox} and Jones\cite{jones}. Cox and Ross\cite{cox} and
Rubinstein\cite{rubensteinone} have solved the problem for the
case when the volatility is a function of the underlying security
price.

Empirical evidence investigating the distribution of stock returns
has shown mixed results. Kon\cite{Kon} finds that the observed
distributions are consistent with stochastic volatility while
Scott\cite{scott} shows that the hypothesis that stock returns are
distributed independently over time can be rejected. Bodurtha and
Courtadon\cite{bodurtha} and Hull and White\cite{hullone} also
support the hypothesis of stochastic volatility. Considering these
results, it seems reasonable to model volatility as another
stochastic variable.

\subsection{The Stochastic Process With Volatility}

Several stochastic processes for the volatility have been
considered by researchers. For example, Hull and
White\cite{hull3}, Heston\cite{heston} and others have considered
the process
\begin{equation}
dV = (a+bV)dt + \xi V^{1/2} dz
\end{equation}
where $Q$ is white noise and $V = \sigma^2$. Baaquie\cite{baaq},
Hull and White and others have considered
\begin{equation}
dV = \mu Vdt + \xi V dz
\end{equation}
while Stein and Stein\cite{stein} consider
\begin{equation}
d\sigma = -\delta(\sigma - \theta) dt + k dz \label{Steineq}
\end{equation}
where $\delta$ and $\theta$ are constants representing the mean
reversion strength and the mean value of the volatility
respectively. We see that all the processes above except for
(\ref{Steineq})\footnote{We can include this process if we add a
term of the form $\gamma V^{1/2}$ to the drift term} follow the
general form
\begin{equation}
dV = (\lambda+\mu V)dt + \xi V^\alpha dz \label{sprocess}
\end{equation} The choice of $\lambda$ and $\mu$ is restricted by
the condition that $V >0$.

\subsection{The Merton-Garman Equation}

The process we are considering is
\begin{align}
dS &= \phi S dt + \sigma S dz_1 \label{proc1}\\ dV &= (\lambda  + \mu
V) dt + \xi V^\alpha dz_2 \label{proc2}
\end{align}
where $\phi,\, \lambda,\, \mu$ and $\xi$ are constants, $V =
\sigma^2$ and $z_1$ and $z_2$ are Wiener processes with
correlation $-1 \le \rho \le 1$. Using Ito's lemma, we obtain the
following expression for the process followed by a derivative
$f_i$ dependent on the underlying security and the volatility of
that security
\begin{equation}
\begin{split}
df_i = &\left(\pdif{f_i}{t} + \phi S\pdif{f_i}{S} + (\lambda + \mu
V)\pdif{f_i}{V} + \frac{\sigma^2 S^2}{2}\pdiftwo{f_i}{S} + \rho
V^{1/2+\alpha} \xi \frac{\partial^2 f_i}{\partial S \partial V} +
\frac{\xi^2 V^{2\alpha}}{2}\pdiftwo{f_i}{V}\right) dt \\ &+
\sigma S\pdif{f_i}{S} dz_1 + \xi V^\alpha \pdif{f_i}{V} dz_2\\
= & \Theta_i dt + \Xi_i dz_1 + \Psi_i dz_2
\end{split}
\end{equation}
We write it in this form to separate the stochastic and non-stochastic
terms.

We now consider two different options, $f_1$ and $f_2$ on the same
underlying security with strike prices and maturities given by
$K_1,\, K_2,\, T_1$ and $T_2$ respectively. We form a portfolio
\begin{equation}
\Pi = f_1 + \Gamma_1 f_2 + \Gamma_2 S
\end{equation}
so that
\begin{equation}
d\Pi = (\Theta_1 + \Gamma_1 \Theta_2 + \Gamma_2 \phi S)dt + (\Xi_1
+ \Gamma_1\Xi_2 + \Gamma_2 \sigma S)dz_1 + (\Psi_1 + \Gamma_1
\Psi_2)dz_2
\end{equation}
We have to get rid of the stochastic terms to ensure perfect
hedging. Hence, we set
\begin{align}
\Xi_1 + \Gamma_1 \Xi_2 + \Gamma_2\sigma S &= 0\\ \Psi_1 + \Gamma_1
\Psi_2 &= 0
\end{align}
to obtain
\begin{align}
\Gamma_1 &= -\frac{\Psi_1}{\Psi_2} = -\frac{\partial f_1/\partial
V}{\partial f_2/\partial V}\\ \Gamma_2 &=
\frac{\Psi_1}{\Psi_2}\pdif{f_2}{S} - \pdif{f_1}{S} =
\frac{\partial f_1/\partial V}{\partial f_2/\partial
V}\pdif{f_2}{S} - \pdif{f_1}{S}
\end{align}

Since the portfolio is now risk-less, it must increase at the
risk-free interest rate by the principle of no arbitrage. In other
words, we must have
\begin{equation}
d\Pi = r\Pi dt
\end{equation}
Expanding $\Pi$ and simplifying, we obtain, after a separation of
variables
\begin{equation}
\begin{split}
 \frac{1}{\partial f_1/\partial V} &\left(
\pdif{f_1}{t} + (\lambda + \mu V) \pdif{f_1}{V} + rS\pdif{f_1}{S}
+ \frac{V S^2}{2} \pdiftwo{f_1}{S} \right.\\ & \left. \quad + \rho
V^{1/2+\alpha} \xi \frac{\partial^2 f_1}{\partial S
\partial V} + \frac{\xi^2 V^{2\alpha}}{2} \pdiftwo{f_1}{V} - rf_1
\right)\\ =\frac{1}{\partial f_2/\partial V} &\left( \pdif{f_2}{t}
+ (\lambda + \mu V) \pdif{f_2}{V} + rS\pdif{f_2}{S} + \frac{V
S^2}{2} \pdiftwo{f_2}{S} \right.\\ &\left. \quad + \rho
V^{1/2+\alpha} \xi \frac{\partial^2 f_2}{\partial S \partial V} +
\frac{\xi^2 V^{2\alpha}}{2} \pdiftwo{f_2}{V} - rf_2 \right)\equiv
\beta(S, V, t, r)
\end{split}
\end{equation}
It is important to note that $\beta$ is not a function of $K_1,\,
K_2,\, T_1$ or $T_2$. This follows from the fact that the first
expression is dependent only on $K_1$ and $T_1$ while the second
is dependent only on $K_2$ and $T_2$. Hence, it is independent of
all four variables. The term $\beta$ is referred to as the market
price of volatility risk. This is because the higher the value of
$\beta$, the more averse the investors are to take on the
volatility risk. The reason this parameter is needed to price
options with stochastic volatility and not for Black-Scholes
pricing is that volatility is not traded in the market. Hence, it
is not possible to perfectly hedge against the volatility even
though it is possible to perfectly hedge against the underlying
security price. Hence, investor risk preferences have to be
considered when considering stochastic volatility or, in other
words, risk-neutral valuation cannot be applied directly to
volatility since volatility is not directly traded in the market.

The parameter, $\beta$, is difficult to estimate empirically and
there is some evidence that it is non-zero\cite{lamoureux}. To
estimate this quantity, we consider the Cox, Ingersoll and Ross
model where the consumption growth has constant correlation with
the spot-asset return. This gives rise to a risk premium which is
proportional to the volatility. We assume this model for
simplicity as it has only the effect of redefining $\mu$ in the
above equation. Henceforth, we shall assume that the market price
of risk has been included in the Merton-Garman equation by
redefining $\mu$. Therefore, the Merton-Garman equation for the
process we are considering is
\begin{equation}
\pdif{f}{t} + rS\pdif{f}{S} + (\lambda + \mu V)\pdif{f}{V} +
\frac{1}{2}VS^2\pdiftwo{f}{S} + \rho \xi
V^{1/2+\alpha}S\frac{\partial^2 f}{\partial S \partial V} +
\xi^2V^{2\alpha}\pdiftwo{f}{V} = rf
\end{equation}
We introduce the variables $S = e^x$ and  $V  =  e^y$  to simplify the
calculations. In terms of these variables, the Merton-Garman
equation is
\begin{equation}\label{mgeq}
\begin{split}
&\pdif{f}{t} + \left(r-\frac{e^y}{2}\right)\pdif{f}{x} +
\left(\lambda e^{-y} + \mu -
\frac{\xi^2}{2}e^{2y(\alpha-1)}\right)\pdif{f}{y} +
\frac{e^y}{2}\pdiftwo{f}{x} + \rho \xi e^{y(\alpha -
1/2)}\frac{\partial^2 f}{\partial x \partial y} \\ &+ \xi^2
e^{2y(\alpha - 1)}\pdiftwo{f}{y} = rf
\end{split}
\end{equation}
For an option, we have $f(T) = \max(e^x-K, 0)$, $T$ being the
maturity time. Hence, this is a final value problem.

\subsection{The ``Straightforward'' Solution when $\boldsymbol{\rho
= 0}$}

When $\rho = 0$, the solution for any volatility process,
stochastic or non-stochastic is straightforward. We make use of
the theorem of Merton that the solution for a deterministic
volatility process is the Black-Scholes price with the volatility
variable replaced by the average volatility. We can consider the
stochastic volatility case as a collection of a large number of
deterministic volatility processes and the option price is then
the average of the prices produced by each of the processes. In
other words, if the volatility follows the generic process $V(t)$
(where $V$ may be stochastic), the option price will be given by
\begin{equation}
C = \int_0^\infty [SN(d_1(V)) - Ke^{-r\tau}N(d_2(V))]V_m(V)dV
\label{rhozero}
\end{equation}
where $V_m$ is the probability distribution function for
the mean of the volatility $\frac{1}{T} \int_0^T V(t) dt$ (which is a delta function for a
deterministic process) and $d_1(V)$ and $d_2(V)$ are the usual variate
in Black-Scholes equation defined by $\dis d_j = \frac{\ln(S/X) + (r
+ (-1)^{j}/2)(T- t)}{\sigma \sqrt{T - t}}, j = 1,2$, $K$ is the
strike price. Here, $N(x)$ is the cumulative probability
distribution function for a standardized normal variate.

This intuitive result is derived in Scott\cite{scott}.

We will give two simple examples to illustrate this. First, let us
consider a deterministic process. We will choose the process
\begin{equation}
V = V_0e^{\mu t},\, 0 \le t \le T
\end{equation}
In this case, the probability distribution function of the mean of
the volatility is given by
\begin{equation}
V_m = \delta\left(V - V_0\frac{e^{\mu T}-1}{\mu T} \right)
\end{equation}
giving us the Black-Scholes result with $\sigma$ replaced by
$\sqrt{V_0\frac{e^{\mu T}-1}{\mu T}}$.

For a stochastic volatility process, we choose\footnote{This is
not a realistic process as $P(V<0)>0$ while $V$ is obviously
non-negative. However, it might be a reasonable approximation for
relatively short times for which $P(V<0)$ is negligible.}$\lambda
= \mu = \alpha =0$ in eq(\ref{proc2}) to obtain
\begin{equation}
dV = \xi dz,\, V(0) = V_0,\, 0 \le t \le T
\end{equation}
where $Q$ represents white noise. The distribution of the mean of
$V$ during the time interval $(0, T)$ is given by
\begin{equation}
V_m \sim N\left(V_0, \frac{\xi^2 T}{3}\right)
\end{equation}
Hence, the option price is given by
\begin{equation}
C = \sqrt{\frac{3}{2\pi \xi^2 T}} \int_0^\infty [SN(d_1(V)) -
Ke^{-r\tau}N(d_2(V))]\exp\left(-\frac{3(V - V_0)^2}{2\xi^2
T}\right) dV.
\end{equation}

\subsection{An Extension to Merton's Theorem} The case $\rho=0$
corresponds to Merton's Theorem. We now extend Merton's theorem to
the case of non-zero correlation for the stochastic process of the
volatility that we are investigating. The present value of the
option is given by
\begin{equation}
f(x, y, T) = \int_{-\infty}^\infty \matel{x,
y}{e^{-\hat{H}\tau}}{x'} f(x', T) dx'
\end{equation}
and $\matel{x, y}{e^{-\hat{H}\tau}}{x'}$ is given by equation
(\ref{propagator}) with $S_0$ and $S_1$ given by (\ref{szero}) and
(\ref{Beautiful}) respectively. Now, since $\int DY e^{S_0}$
describes the probability of a specific path for $y$ (we show this
in detail in chapter 6), we see that the propagator can now be
written as in equation (\ref{Beautifulone}) with $\omega,\,
\theta,\, \eta$ and $\zeta$ being functionals of this path and
$\nu$ being the final value of $V^{3/2-\alpha}$ for the path.
Hence, if we generate paths for $y$ according to (\ref{proc2}),
the option price is given by
\begin{equation}
f(x, y, t) = \left\langle \int_{\ln K}^\infty dx'
\frac{e^{S_1(x,\, x', \, \omega,\, \theta,\, \eta,\, \zeta,\,
\nu)}}{\sqrt{2\pi \epsilon (1-\rho^2)}} (e^{x'}-K)_+ \right\rangle
\label{prop1}
\end{equation}
(since the payoff of the option is given by $(e^{x'}-K)_+$) with
the average taken over the paths for $V$. Since the propagator is
in the form of a Gaussian, we can perform the integration over
$x^\prime$ to obtain
\begin{equation}
\begin{split}
f(x, y, t) = &\left\langle SN(s_1)\exp\left(
-\frac{\rho^2}{2}\omega - \frac{\rho}{(3/2-\alpha)\xi}\left(
V_i^{3/2-\alpha}-V_f^{3/2-\alpha} \right) -
\frac{\rho\lambda}{\xi} \theta - \frac{\rho\mu}{\xi} \eta \right.
\right.\\ & \quad\left.  \left.  +\frac{\rho\xi}{2} \zeta \right)
- Ke^{-r\tau}N(s_2) \right\rangle \\
\end{split}
\label{SrikantBaaquie}
\end{equation}
where $V_i$ and $V_f$ are the initial and final volatilities of
the path respectively and $s_1$ and $s_2$ are given by
\begin{align}
s_1 = &\frac{\ln\left(\frac{S}{K}\right) + r\tau + \frac{1}{2}
(1-2\rho^2) \omega + \frac{\rho}{(3/2-\alpha)\xi}
\left(V_i^{3/2-\alpha} - V_f^{3/2-\alpha} \right) +
\frac{\rho\lambda}{\xi} \theta + \frac{\rho\mu}{\xi} \eta -
\frac{\rho\xi}{2} \zeta}{\sqrt{(1-\rho^2)\omega}}\\ s_2 = &s_1 -
\sqrt{(1-\rho^2)\omega}
\end{align}
It is easy to verify that equation (\ref{SrikantBaaquie}) is the
same as the Black-Scholes equation for any single volatility path
when $\rho = 0$.

When $\alpha = \frac{1}{2}$, several simplifications occur. We see
that $\theta = \zeta = \tau$ are known and $\eta = \omega$. In
that case, (\ref{SrikantBaaquie}) reduces to
\begin{equation}
 \left\langle SN(s_1)\exp\left( -\left(\frac{\rho^2}{2} +
\frac{\rho\mu}{\xi}\right) \omega - \frac{\rho}{\xi}
\left(V_i-V_f\right) - \rho\left(\frac{2\lambda -
\xi^2}{2\xi}\right) \tau \right) - Ke^{-r\tau}N(s_2) \right\rangle
\label{eqhalf}
\end{equation}
where $s_1$ and $s_2$ now have the relatively simple forms
\begin{align}
s_1 = &\frac{\ln\left(\frac{S}{K}\right) + \left(r + \rho\left(
\frac{2\lambda - \xi^2}{2\xi}\right) \right) \tau - \left( \rho^2
- \frac{\rho\mu}{\xi} - \frac{1}{2} \right) \omega +
\frac{\rho}{\xi} \left(V_i-V_f\right)}{\sqrt{(1-\rho^2)\omega}}\\
s_2 = &s_1 - \sqrt{(1-\rho^2)\omega}
\end{align}
Hence, we see that we have a straightforward solution for $\alpha
= \frac{1}{2}$ even when the correlation is not zero.

When $\alpha = \frac{3}{2}$ and $\lambda = 0$, we obtain a similar
simplification since $\zeta = \omega$ and $\eta = \tau$. In this
case, we obtain the following expression for the option price
\begin{equation}
f(x, y, t) = \left\langle SN(s_1)\exp\left(-\frac{\rho}{2} \left(
(\rho - \xi) \omega + 2\ln\left(\frac{V_i}{V_f}\right) + 2
\frac{\mu\tau}{\xi}\right) \right) - Ke^{-r\tau}N(s_2)
\right\rangle
\end{equation}
and $s_1$ and $s_2$ are now given by
\begin{align}
s_1 = &\frac{\ln\left(\frac{S}{K}\right) + \left(r +
\frac{\rho\mu}{\xi} \right) \tau + \left( -\rho^2 -
\frac{\rho\xi}{2} + \frac{1}{2} \right) \omega + \frac{\rho}{\xi}
\ln\left( \frac{V_i}{V_f} \right)}{\sqrt{(1-\rho^2)\omega}}\\ s_2
= &s_1 - \sqrt{(1-\rho^2)\omega}
\end{align}

For the case considered in Baaquie\cite{baaq}, we have $\lambda =
0$ and $\alpha = 1$. In this case, we have $\eta = \zeta =
\int_0^\tau e^{y/2} dt$ which gives us
\begin{equation}
f(x, y, t) = \left\langle SN(s_1)\exp\left(
-\frac{\rho^2\omega}{2} - \frac{2\rho}{\xi}\left( V_i^{1/2} -
V_f^{1/2} \right) - \frac{\rho}{\xi} \left( \mu - \frac{\xi^2}{2}
\right) \eta \right) - Ke^{-r\tau}N(s_2) \right\rangle
\end{equation}
where $s_1$ and $s_2$ are now given by
\begin{align}
s_1 = &\frac{\ln\left(\frac{S}{K}\right) + r\tau + \left( -\rho^2
+ \frac{1}{2} \right) \omega + \frac{2\rho}{\xi} \left( V_i^{1/2}
- V_f^{1/2} \right) + \frac{\rho}{\xi}\left( \mu - \frac{\xi^2}{2}
\right) \eta}{\sqrt{(1-\rho^2) \omega}}\\ s_2 = &s_1 -
\sqrt{(1-\rho^2) \omega}
\end{align}
which is somewhat more complicated since two functionals, $\omega$
and $\eta$ of the volatility path are involved. In this case,
however, a perturbation  analysis can be used to derive an
approximate form for the probability distribution functions of the
functionals. Due to this fortunate occurrence, a series solution
to this problem can be obtained.

The probability density function (pdf) for the functionals is a
very difficult quantity to obtain. The probability density
function for $\int_0^\tau Vdt$ was obtained for the special case
$\alpha = 1/2$ in Stein and Stein\cite{stein}\footnote{The
original solution for simple Brownian motion was obtained way back
in 1944 by Cameron and Martin\cite{cameron}}. Stein and
Stein\cite{stein} have used this probability distribution function
and the ``straightforward'' solution for $\rho = 0$ to get an
analytic form of the solution for this case. We now see that the
result can be extended to non-zero $\rho$ if we can find the joint
probability density function of this functional and  $V_i-V_f$.
While the individual probability distribution functions can be
obtained (the pdf for $\omega$ is obtained in Stein\cite{stein}
and the pdf for $V_i-V_f$ is trivial), they are not independent.

\subsection{Risk-Neutrality} We show that the expected value of the
underlying security $S$ whose initial value is $S_0$ is given by
$S_0e^{rt}$ after time $t$ has elapsed for a large class of
stochastic processes including the one we are considering in this
thesis. In other words, we show that $A = e^{-rt}S$ is a
martingale.

We first change variables from $S$ to $A$ in (\ref{proc1})
(changing $\phi$ to $r$ in accordance with risk-neutral valuation)
to obtain
\begin{equation}
dA = Ae^{y/2} dz_1
\end{equation}
(where $z_1$ is the time integral of $W$ and hence a Wiener
process) where $y$ may depend on $A$ ($y$ can be stochastic).  We
now consider the more general process
\begin{equation}
dA = f(A, y) dz_1
\end{equation}
We note that $E[dA] = 0$ so that $E[A(t+dt)|A(t) = A_0] = A_0$. In
other words, we see that $A$ is a martingale. Hence, we have shown
the result. In general, a martingale process cannot have a drift
term.

While the result is simple, it has important consequences. We note
that risk-neutrality alone cannot determine any constraints for
the volatility process. Any volatility process whatsoever
satisfies risk-neutrality.

%%%%%%%%%%%%%%%%%%%%%%%%%%%%%%%%%%%%%%%%%%%%%%%%%%%%%%%%%

\section{Numerical Algorithm}\label{generalalgo}
\begin{comment}
The bisection method mentioned in the previous section does not
generally work for the case in which $\alpha \neq 1$ and $\lambda
\neq =0$. In the general case, we can still perform the functional
integration easily using the Monte Carlo method.
\end{comment}

\subsection{A Short, Quick Reminder of the Monte Carlo Method} The
Monte Carlo algorithm to integrate
\begin{equation}
\int_A f(X) dX
\end{equation}
where $X$ can be (and, in fact, usually is) a multi-dimensional
variable and $A$ is a subset of the domain of $X$ requires us to
split $f(X) = g(X) p(X)$ so that $\int_A p(X)dX = 1$ (in other
words, so that $p(X)$ is a valid probability density function).
The algorithm then states that an estimate for the integral is
given by $\langle g(X_i) \rangle = \frac{1}{N} \sum_{i=1}^N
g(X_i)$ where the configurations $X_i$ are generated randomly
according to the probability density function $p(X)$.

The error of the Monte Carlo method goes as $\frac{1}{\sqrt{N}}$
as a consequence of the central limit theorem as long as there is
no correlation between the configurations produced. (Though in
general this condition is difficult to satisfy, we shall see later
that we can easily satisfy it for this case). While this error may
not look very impressive, it is often the best that can be managed
for $X$ which have a large number of dimensions. For the present
problem, we have a very large number of dimensions (in fact the
exact problem has an infinite number of dimensions) and the
Monte-Carlo method is the most practical one available for it.

\subsection{A Monte Carlo Method for this Problem}
The Monte Carlo based numerical approaches of Hull and White
\cite{hull2,hull3}, Finucane \cite{fin} and Mills \cite{mill} are
all generalized forms of the Binomial tree in which for discrete
time $t_n = n\frac{\tau}{N}$, the stock price $x_n$ and volatility
$y_n$ are considered as random variables and both of which are
updated.  For the case of $N$-steps, we need to update $N^2$
variables for obtaining a new configuration.  In
eq(\ref{propagator}), we have a drastic simplification since the
path integral over the $x_i$ variables for the general case of
$\rho \neq 0$ has been evaluated {\it exactly} by analytical
means. Hence in basing numerical simulation on
eq(\ref{propagator}), one needs to generate new configurations for
only the $y_n$ random variables, namely $N$ variables, reducing
the computational time required by a factor of $N$.

For this problem, we choose the following probability density
function
\begin{equation}
p(Y) = \left( \prod_{i=1}^{N-1}
\frac{e^{y_i(1-\alpha)}}{\sqrt{2\pi\epsilon} \xi} \right) e^{S_0}
\label{probY}
\end{equation}
where $Y$ is the set of variables $y_i$ (and is hence
$N$\/-dimensional) and $S_0$ is given in (\ref{szero}). Hence, we
see that
\begin{equation}
g(Y) = \frac{e^{S_1}}{\sqrt{2\pi\epsilon (1-\rho^2) \sum_{i=1}^N
e^{y_i}}} \label{defineg}
\end{equation}
where $S_1$ is given in (\ref{sone}) since the integral we are
performing is
\begin{equation}
\int DY \left( \prod_{i=1}^{N-1} e^{y_i(1-\alpha)} \right)
\frac{e^{S_0+S_1}}{\sqrt{2\pi \epsilon (1-\rho^2) \sum_{i=1}^N
e^{y_i}}}
\end{equation}
where
\begin{equation*}
DY = dy_0 \prod_{i=1}^{N-1} \frac{dy_i}{\xi \sqrt{2\pi\epsilon}}
\end{equation*}

We now have to produce configurations $Y$ with the probability
distribution $p(Y)$. While $p(Y)$ looks rather complicated, it has
a simple interpretation. It is the probability distribution for a
discretized random walk performed by $y$. To see this, let us
first use Ito's lemma to find the process followed by $y$. We
find, from eq(proc2) and eq(variable),
\begin{equation}
dy = \left( \lambda e^{-y} + \mu - \frac{\xi^2
e^{2y(\alpha-1)}}{2} \right)dt + \xi e^{y(\alpha-1)} Q dt
\label{yprocess}
\end{equation}
We can now discretize the process using Euler's method to obtain
\begin{equation}
\delta y_i = \left( \lambda e^{-y_i} + \mu - \frac{\xi^2
e^{2y_i(\alpha-1)}}{2}\right) \epsilon + \xi e^{y_i(\alpha-1)} Z
\sqrt{\epsilon}
\end{equation}
where $\delta y_i = y_{i} - y_{i-1}$, $\epsilon$ is the time step
and $Z$ is a standard normal variable.  Since we are using $\tau =
T-t$ as the time variable, the time step is actually $-\epsilon$.
Hence, $\delta y_i$ is a normal random variable with mean $\left(
-\lambda e^{-y_i} - \mu +
\xi^2e^{2y_i(\alpha-1)}/2\right)\epsilon$ and variance
$\xi^2e^{2y_i(\alpha-1)}\epsilon$ and the probability density
function is given by
\begin{equation}
f_i = \frac{e^{y_i(1-\alpha)}}{\xi \sqrt{2\pi\epsilon}} \exp\left(
-\frac{\epsilon e^{2y_i(1 - \alpha)}}{2\xi^2} \yexpri^2 \right)
\end{equation}
Hence, the joint probability density function for the discretized
process is given by
\begin{equation}
f = \prod_{i=1}^{N-1} f_i = \left( \prod_{i=1}^{N-1}
\frac{e^{y_i(1-\alpha)}}{\sqrt{2\pi\epsilon}\xi} \right) e^{S_0}
\end{equation}
which is the same as (\ref{probY}).

In this simulation, we will use Euler's method to find the
volatility paths since these paths are generated with the
requisite probability distribution.

The algorithm to find a Monte Carlo estimate of the propagator $p
= \matel{x, y}{e^{-\hat{H}\tau}}{x'}$ is as follows
\begin{enumerate}
\item p := 0 (Initialization)
\item For i := 1 to N
\item Generate a path $Y$ for $y$ using (\ref{yprocess})
\item $p := p+g(Y)/N$ (where $g(Y)$ is defined in (\ref{defineg}))
\item End For
\end{enumerate}
The paths must be generated backwards starting from $y_N$ which is
the initial value of $\ln V$ to obtain all the $y_0$. Since the
equations are time symmetric (after, of course, reversing the
drift terms), this presents no problem. This will have to be
repeated for all the $x'$ that we wish to integrate over. However,
during implementation it is found to be more advantageous to
generate the paths only once, storing the important terms $t = N
\epsilon$, $t_1 = \sum_{i=1}^N e^{y_i}$ and $$t_2
=\frac{\rho}{\epsilon} \sum_{i=1}^N e^{y_i(3/2-\alpha)} \yexpri$$
which are sufficient to determine $S_1$ once $x'$ is given ($S_1 =
-\frac{1}{2\epsilon (1-\rho^2) t_1}\left( x-x' + (r-q)t - \epsilon
\left( \frac{t_1}{2}+t_2 \right) \right)^2$) where $q$ is the
annualized dividend. That $S_1$ can be computed using this limited
information is fortunate as storing all the paths explicitly would
require a very large memory (10MB for 10,000 configurations as
compared to 160kB when storing only the essential combinations of
terms). The alternative of generating paths for each value of $x'$
(as in the na\"ive algorithm above) is unacceptable due to the
very large run time required using this approach.

The propagator must finally be integrated over the final wave
function to obtain the option price. The accuracy of this
numerical quadrature depends on the spacing $h$ between successive
values of $x'$. This means that we have to find the propagator for
several values of $x'$ to obtain reasonable accuracy which is
computationally very expensive. We found it better to find the
propagator using the above Monte Carlo method for only about 100
equally spaced values of $x'$ over the range of the quadrature and
using cubic splines to interpolate it at the other quadrature
points. This produces excellent results (as we shall see later) as
the propagator seems to be an extremely smooth function of $x'$.

Hence, the revised algorithm is of the form
\begin{enumerate}
\item For i := 1 to N
\item Generate a path $Y$ for $y$ using (\ref{yprocess})
\item Store $t_1$ and $t_2$ for the path
\item End For
\item For $x'$ := beginning of range to end of range
\item Find the Monte Carlo estimate for the propagator at large
intervals of $x'$ using $t_1$ and $t_2$ from the paths.
\item End For
\item For $x'$ := beginning of range to end of range
\item Find the propagator at small intervals of $x'$ using cubic
spline interpolation over the values of the propagator found
previously and integrate over the final wave function (payoff)
\item End For
\item Return option price
\end{enumerate}
The numerical quadrature was performed using Simpson's rule. While
Simpson's rule is a relatively low order method, it was deemed
sufficient as the error in quadrature is negligible compared to
the Monte Carlo error in the propagator. For the special case of
the European call option, it is possible to integrate analytically
over the splines as the payoff is piecewise linear. However, this
was not done as firstly, the error in the quadrature was
negligible compared to the Monte Carlo error and secondly (and
more importantly) we wanted to make the program general so that it
is able to price any derivative based on the underlying security
which has only one payoff date.

For the special case of functions which can be expressed as
piecewise polynomial functions, the integration can be carried out
analytically and we can then just generate the paths and find the
expectation over the generalized Black-Scholes prices. While this
method has the advantages of being elegant, easy to program and
completely eliminating the error due to the integration, it has
the disadvantage that we have to carry out the integration first
before we can write the program which will be specific to the
given payoff. We have implemented this approach for call options
 but found no significant
improvement in run time (it runs about 1\% faster).

The above algorithm may appear unstable in the limit $\epsilon
\tendsto 0$ since in this case $\frac{\delta y}{\epsilon} =
O\left(\sqrt{\frac{1}{\epsilon}}\right)$. Fortunately, in the case
of our simulations this is not a problem as our step size was
always large enough to avoid this problem. If this is a problem in
any simulation, this can be easily handled as the propagator can
always be written as a functional of the paths as given in
(\ref{prop1}). We can then generate paths for $V$ or $y$ using
(\ref{proc2}), store the functionals (we will now need 5 terms
$t_1 = \omega$ (which is the same as above), $t_2 = \theta =
\sum_{i=1}^N e^{y_i(1/2-\alpha)}$, $t_3 = \eta = \sum_{i=1}^N
e^{y_i(3/2-\alpha)}$, $t_4 = \zeta = \sum_{i=1}^N e^{y_i(\alpha -
1/2)}$ and $t_5 = e^{y_0(3/2-\alpha)}$). We can then proceed to
find the propagator as above (of course, now writing $S_1$ as a
function of $t_i,\, i = 1,\dots, 5$) and integrate over the final
payoff. For the special case of the European call option, one can
make this method slightly more efficient by averaging over the
generalized Black-Scholes prices (\ref{SrikantBaaquie}). We did
not use this method since it takes longer to compute the
propagator using five terms and because the memory requirements
are higher.
%%%%%%%%%%%%%%%%%%%%%%%%%%%%%%%%%%%%%%%%%%%%%%%%%%%%%%%%%%%%%%%%%%%%%

\section{Numerical Results for the General Case}\label{gensim}

We performed simulations on 90 day options setting an initial
volatility of 25\% per annum which is a figure comparable to the
historical data. We performed simulations for $\alpha = 0,\,
\frac{1}{2},\, \frac{3}{4}$ and 1 with correlations ranging from
-0.5 to 0.5 in steps of 0.1. We set $\lambda = \mu = 0$ for most
of the simulations. Some simulations with mean-reverting
volatility processes were performed for the purposes of
investigating the effect of mean-reversion on option prices.

The following parameters have the following values unless
otherwise stated : $S_0 = 100,\, r = 5\%,\, q = 0\%, \, t = 90
\text{ days},\, \lambda = 0$ and $\mu = 0$. Most of the
simulations have been performed using 128 time steps and 500,000
Monte Carlo configurations. The exceptions have been for $\alpha =
\frac{1}{2}$ where we have used 512 time steps and the simulations
we have performed for the other values of $\alpha$ to check that
the number of time steps is sufficient. The error bars in all the
graphs refer to the standard error obtained from the Monte Carlo
simulation.

\subsection{The Effect of $\rho$ on Option Prices}
\begin{figure}[ht]
\begin{tabular}{cc}
\epsfig{file =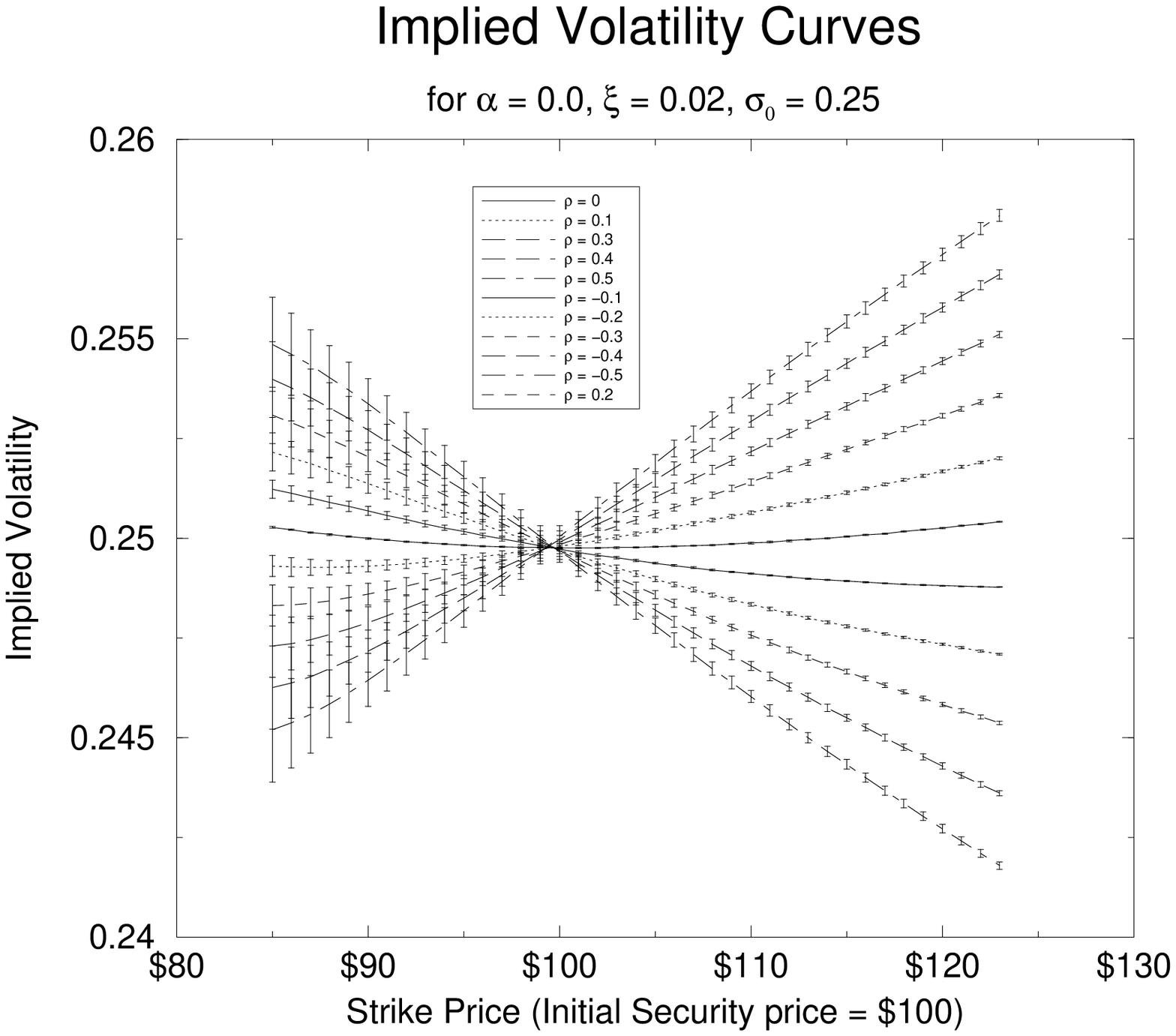, width = 7.0cm, height=8cm} &
\epsfig{file =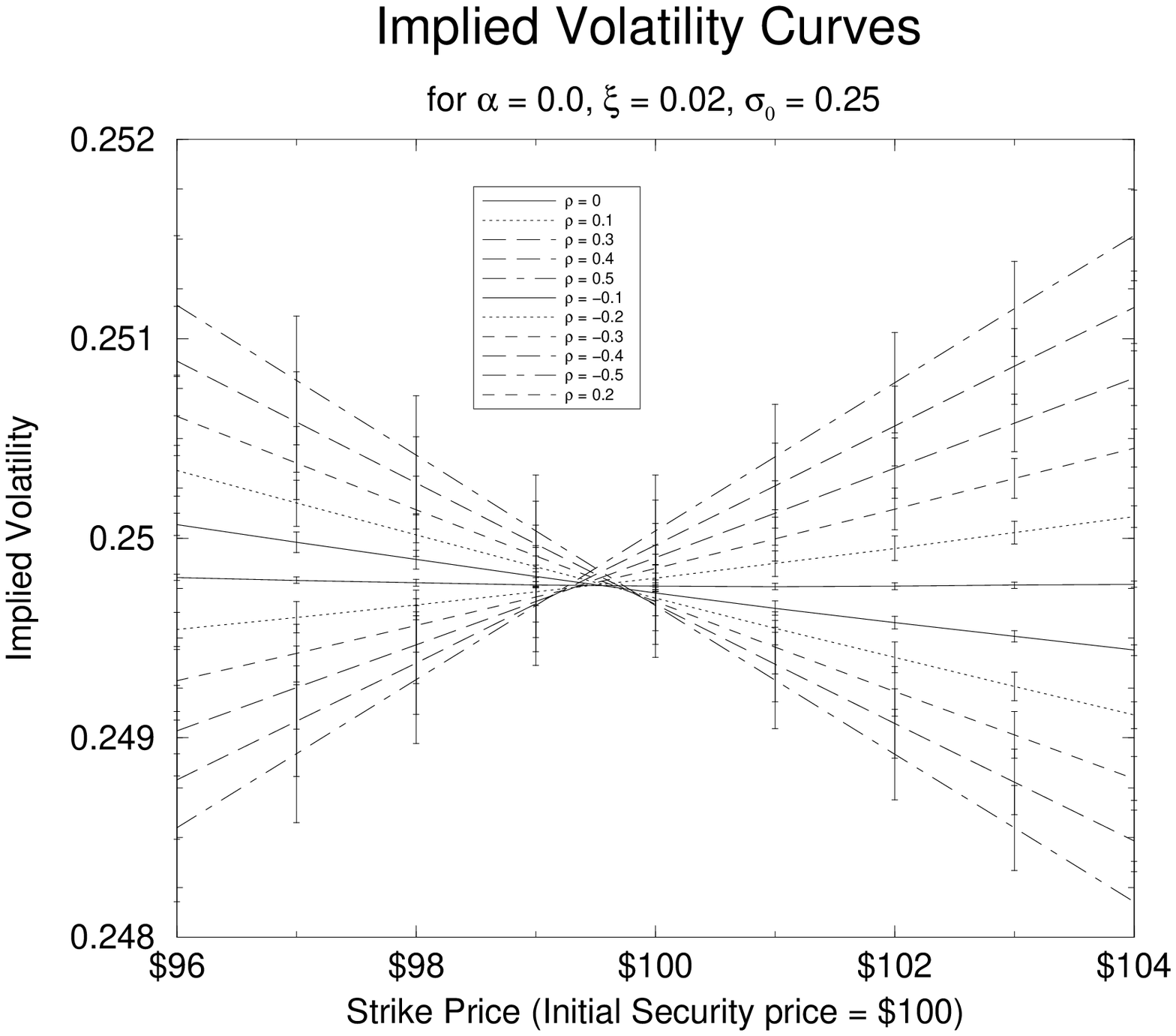, width = 7.0cm, height=8cm}\\
\end{tabular}
%\centerline{graph1a.eps and graph1b.eps}
\caption{Implied volatility curves showing the effect of $\rho$ on
option prices when $\alpha = 0$. We can see that positive $\rho$
leads to an increase in the option price when the strike price is
high and a decrease when the strike price is low while negative
$\rho$ has the opposite effect.} \label{Table_rho1}
\end{figure}

\begin{figure}[ht]
\begin{tabular}{cc}
\epsfig{file =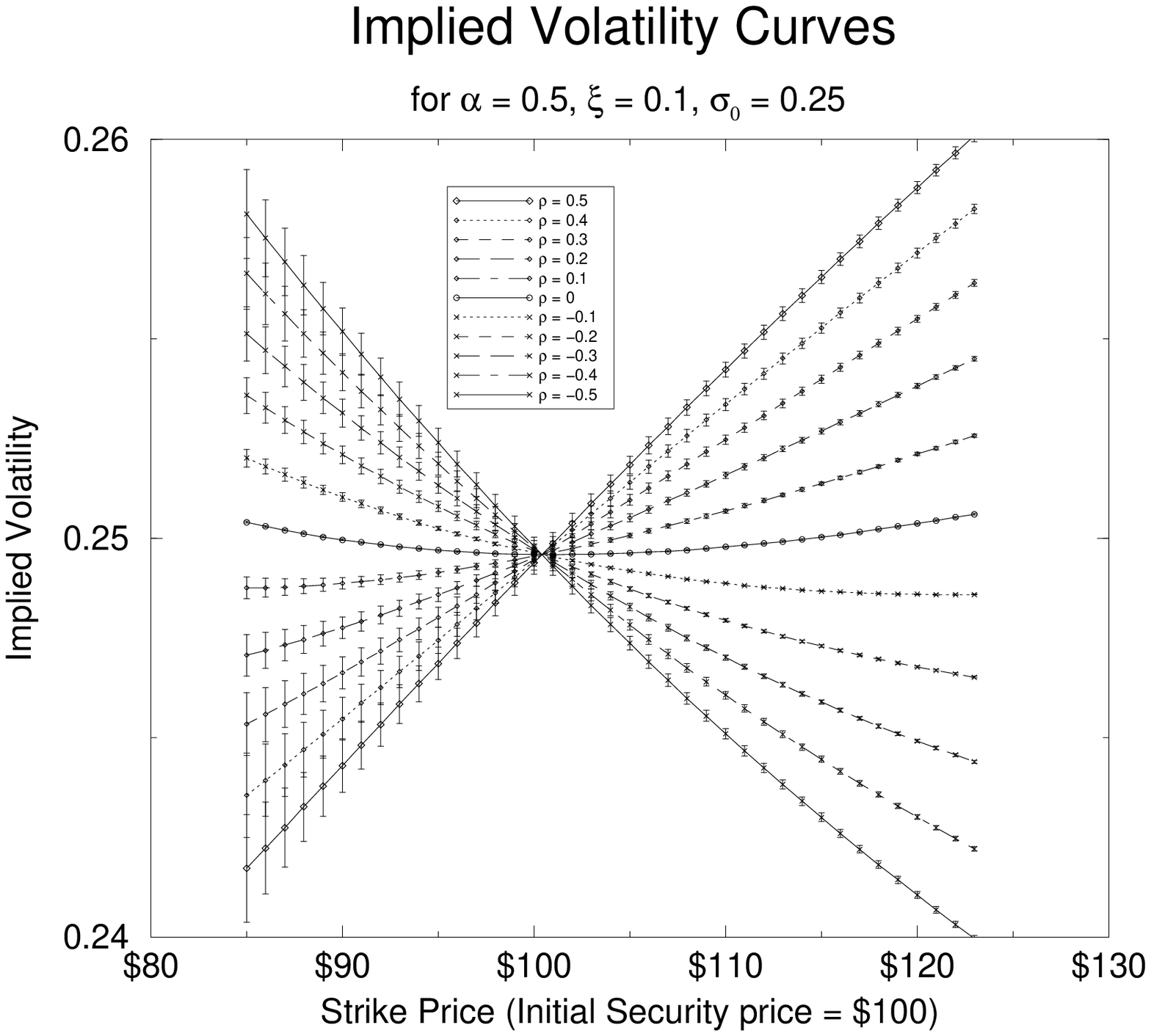, width = 7.0cm, height=8cm} &
\epsfig{file =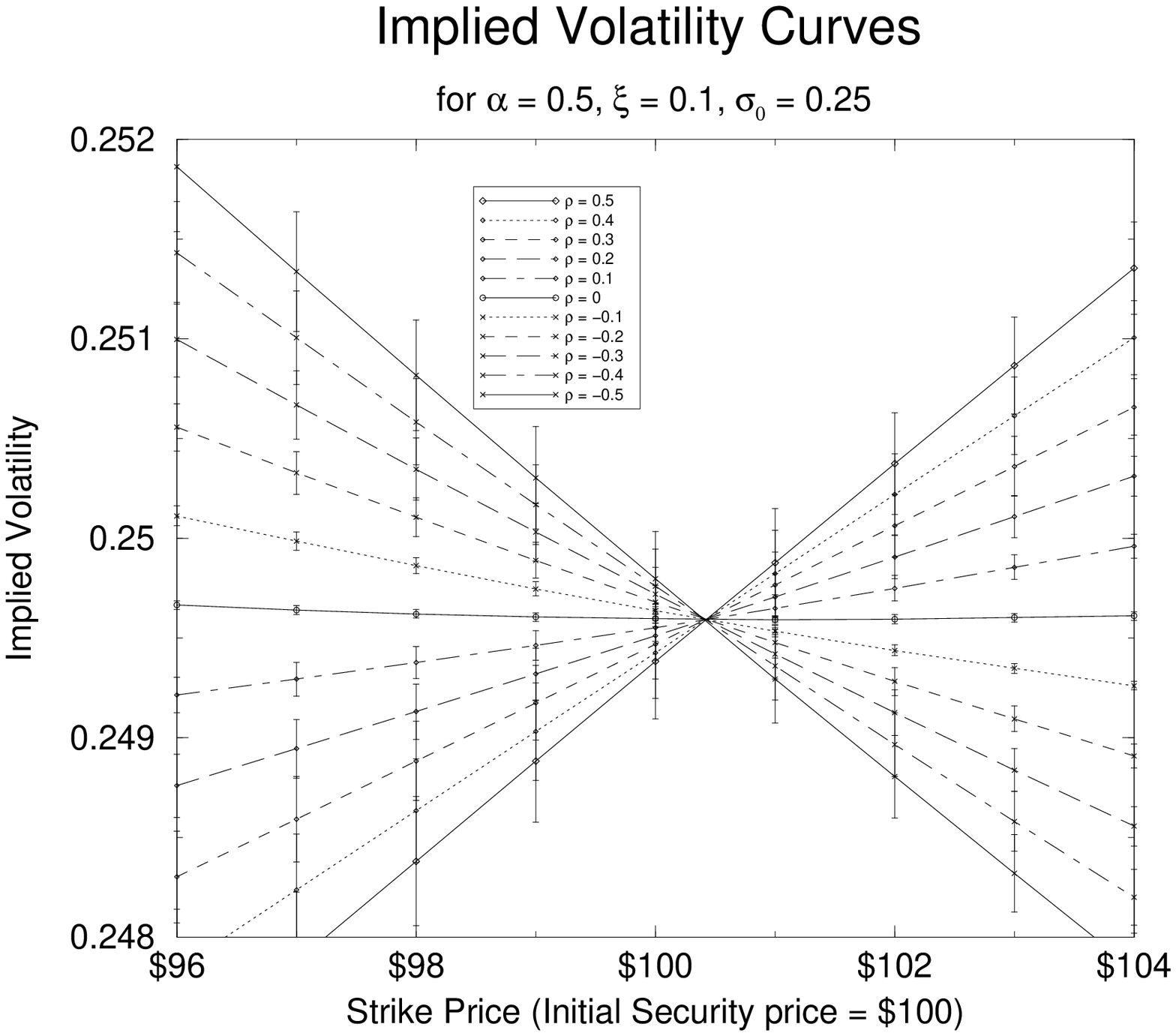, width = 7.0cm, height=8cm}\\
\end{tabular}
%\centerline{graph1b.eps and graph2b.eps}
\caption{Implied volatility curves showing the effect of $\rho$ on
option prices when $\alpha = \frac{1}{2}$. We can see that
positive $\rho$ leads to an increase in the option price when the
strike price is high and a decrease when the strike price is low
while negative $\rho$ has the opposite effect.} \label{Table_rho2}
\end{figure}

\begin{figure}[ht]
\begin{tabular}{cc}
\epsfig{file =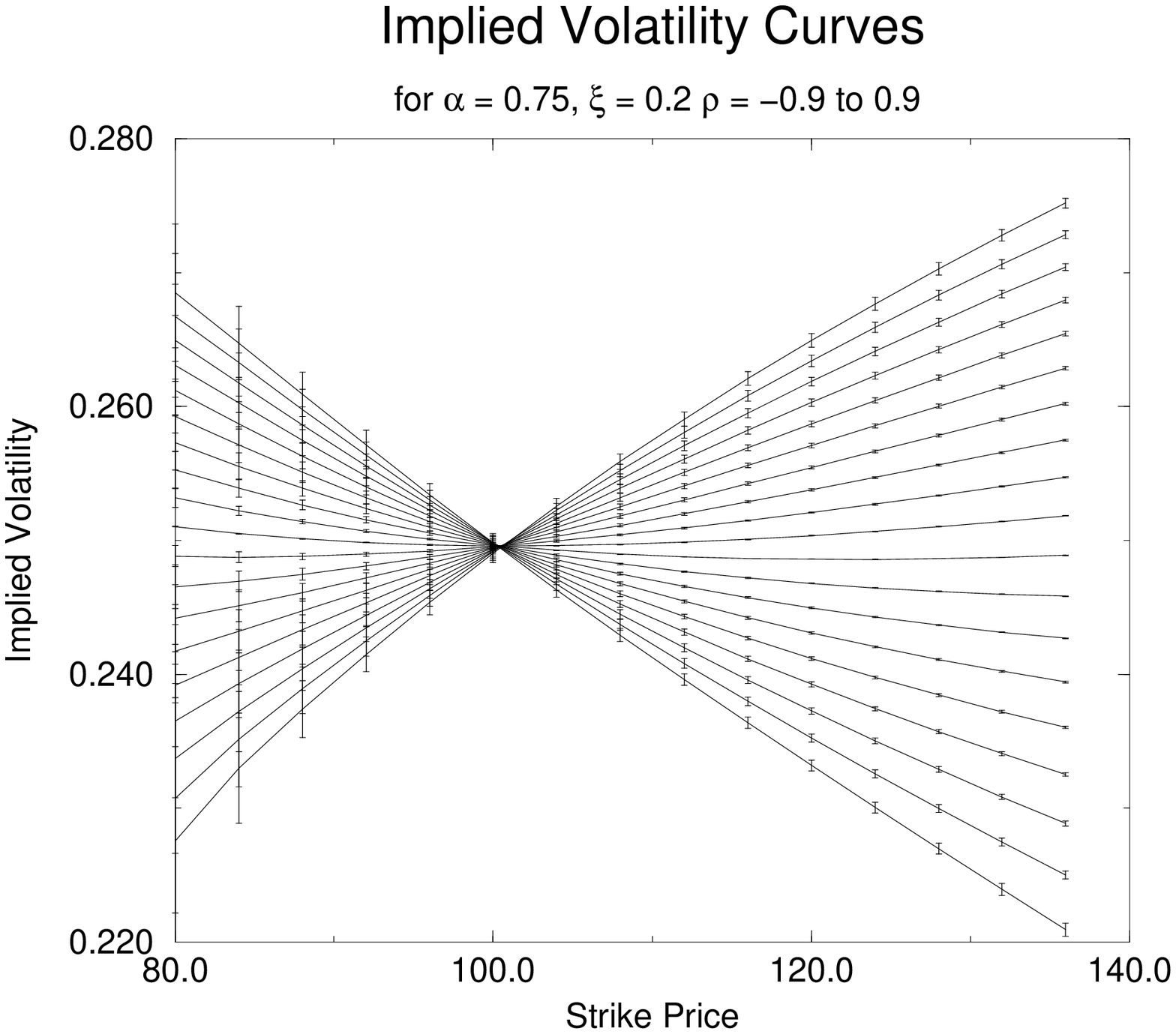, width = 7.5cm, height=8cm} &
\epsfig{file =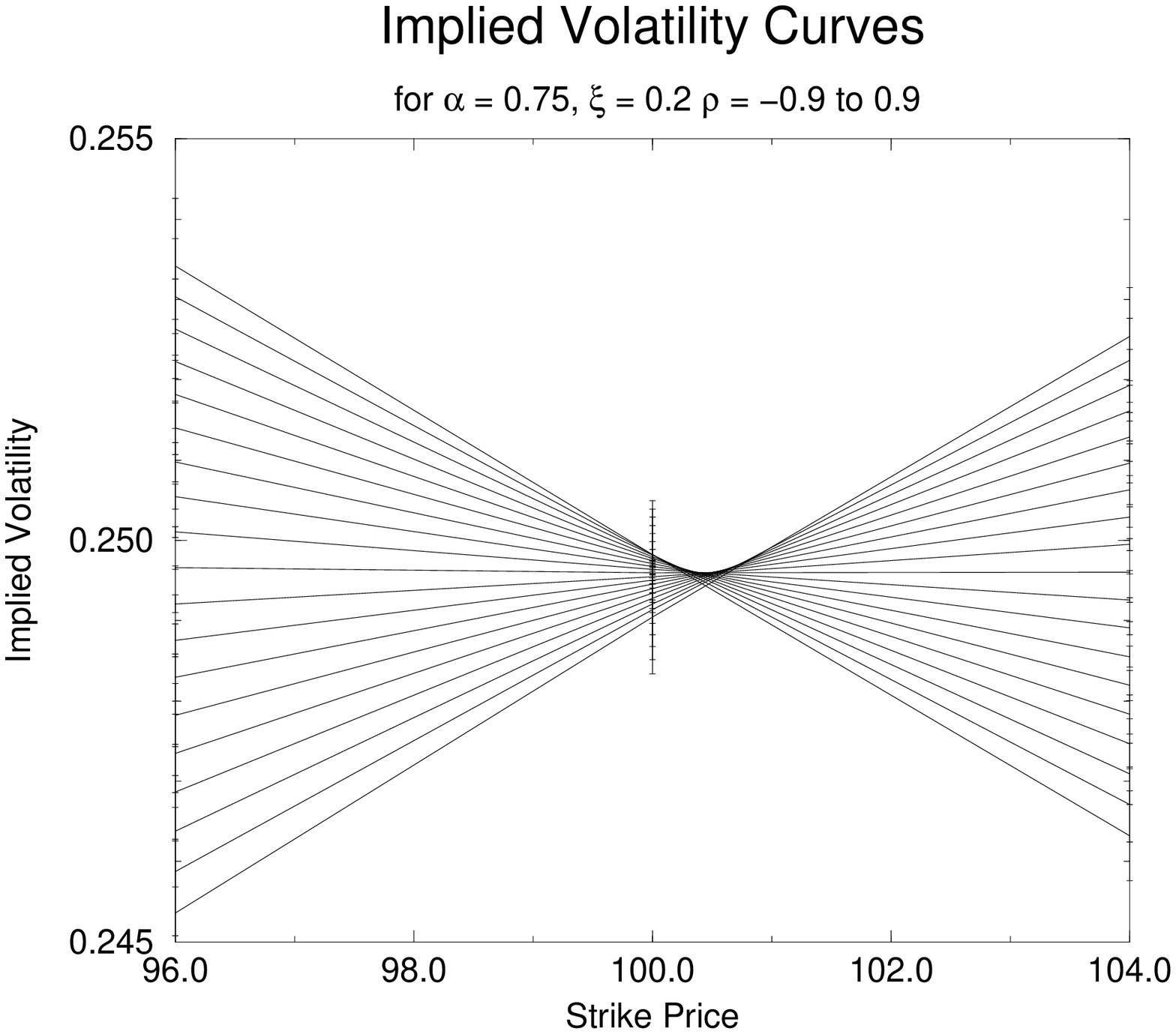, width = 7.5cm, height=8cm}\\
\end{tabular}
%\centerline{graph1d.eps and graph2d.eps}
\caption{Implied volatility curves showing the effect of $\rho$ on
option prices for $\alpha = 1$. We can see that positive $\rho$
leads to an increase in the option price when the strike price is
high and a decrease when the strike price is low while negative
$\rho$ has the opposite effect.} \label{Table_rho3}
\end{figure}

\begin{figure}[ht]
\begin{tabular}{cc}
\epsfig{file =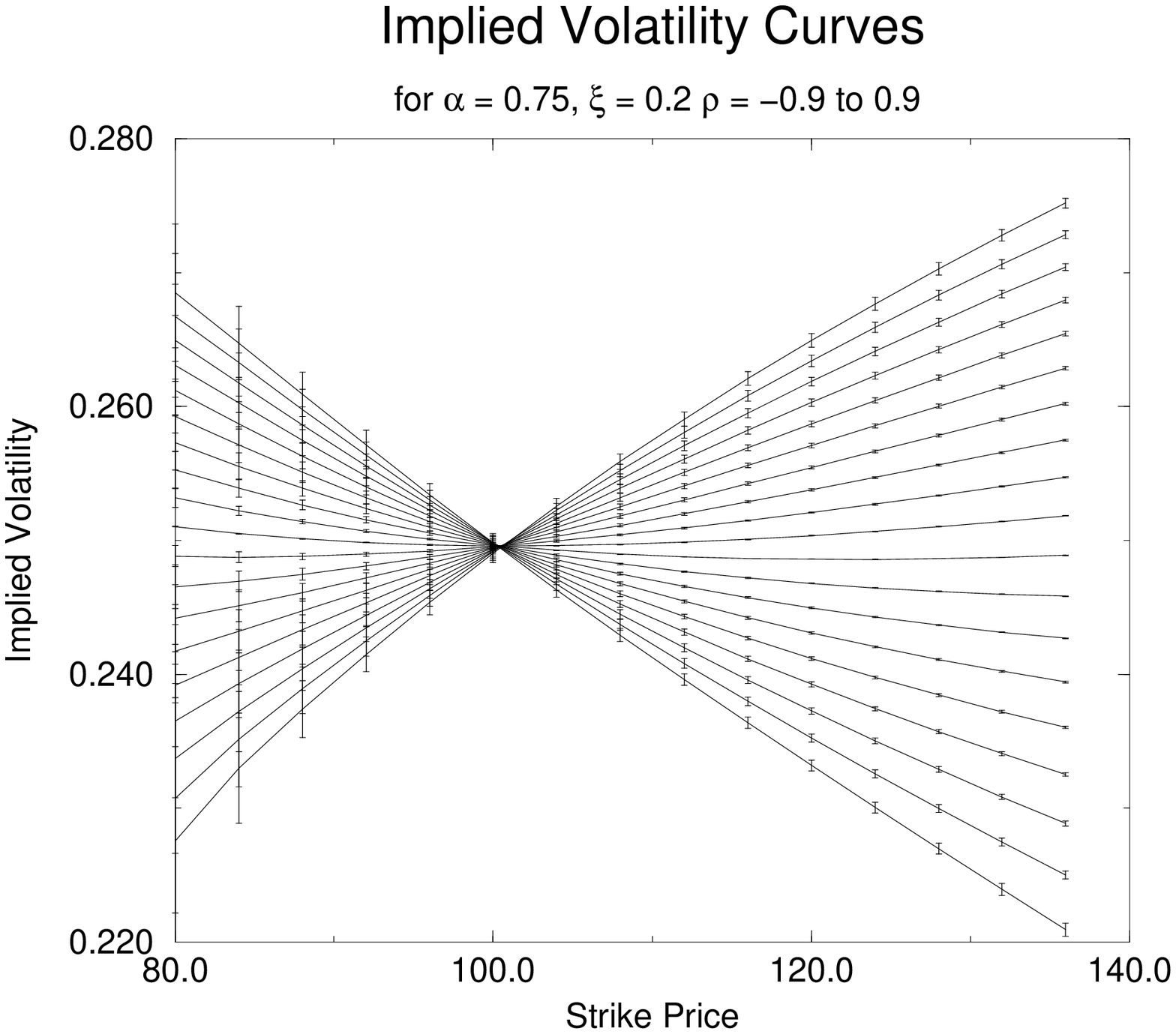, width = 7.5cm, height=8cm} &
\epsfig{file =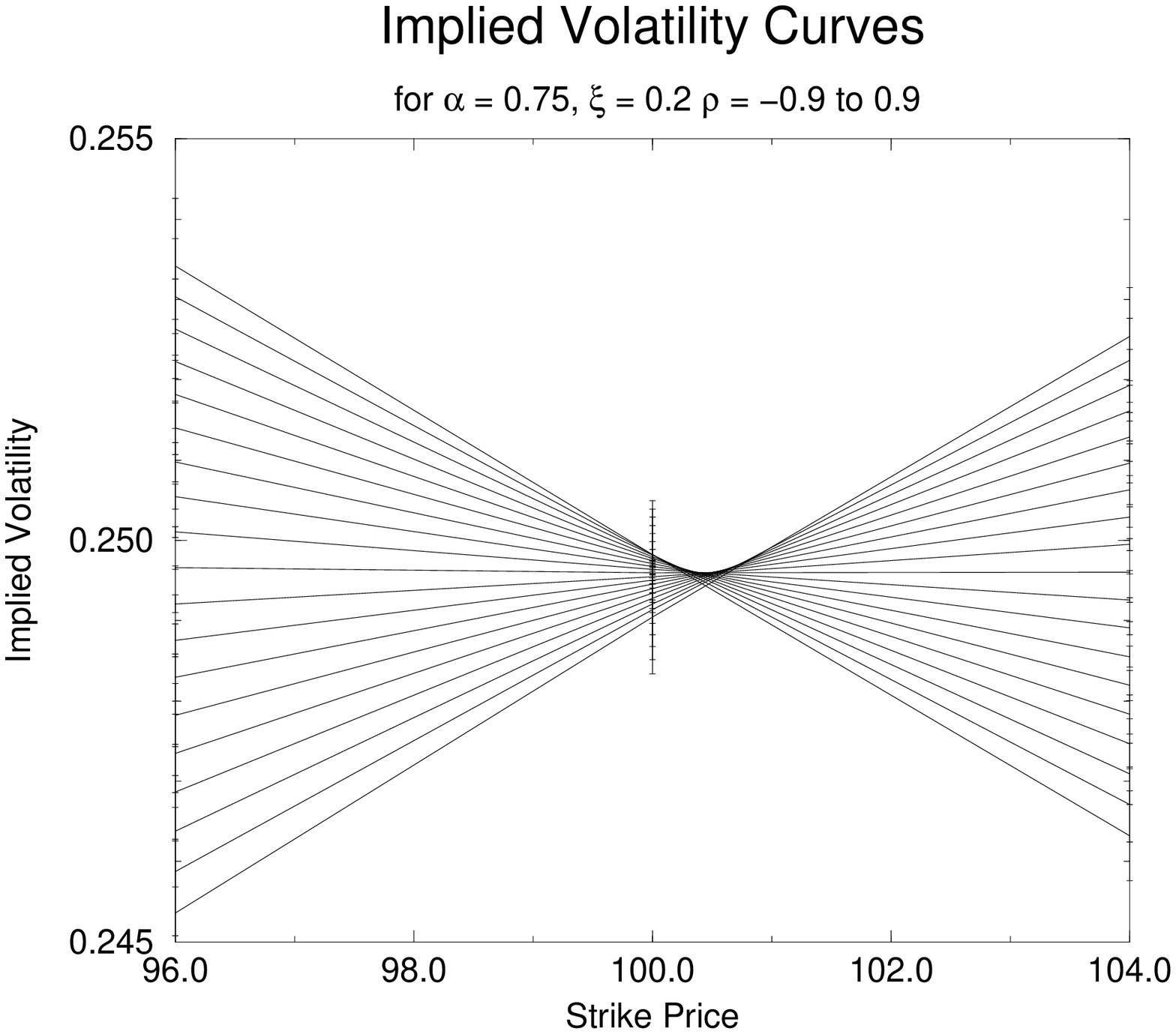, width = 7.5cm, height=8cm}\\
\end{tabular}
%\centerline{graph1e.eps and graph2e.eps}
\caption{Implied volatility curves showing the effect of $\rho$ on
option prices for $\alpha = 1$. The curves for the different
values of $\rho$ are in ascending order according to the slope (in
other words, the slope increases monotonically with $\rho$).
Hence, we can see that positive $\rho$ leads to an increase in the
option price when the strike price is high and a decrease when the
strike price is low while negative $\rho$ has the opposite
effect.} \label{Table_rho4}
\end{figure}

The correlation $\rho$ has a very large impact on the implied
volatility curve irrespective of the values of the other
parameters. When $\rho = 0$, we see that the implied volatility is
in the form of a smile with the minimum near the present value of
the underlying security ($S_0$). As $\rho$ increases, we find that
the implied volatility increases for large strike prices and
decreases for small strike prices. This can be easily seen in the
graphs in figures \ref{Table_rho1}, \ref{Table_rho2},
\ref{Table_rho3} and \ref{Table_rho4}. We can also see that the
deviation of the implied volatility from the initial volatility is
much higher when the correlation is non-zero. These results are
consistent with those reported in Hull and White\cite{hullone},
Heston\cite{heston}, Johnson and Shanno\cite{johnson} and
Scott\cite{scott}.

\begin{figure}[ht]
\begin{center}
\begin{tabular}{c}
\epsfig{file = 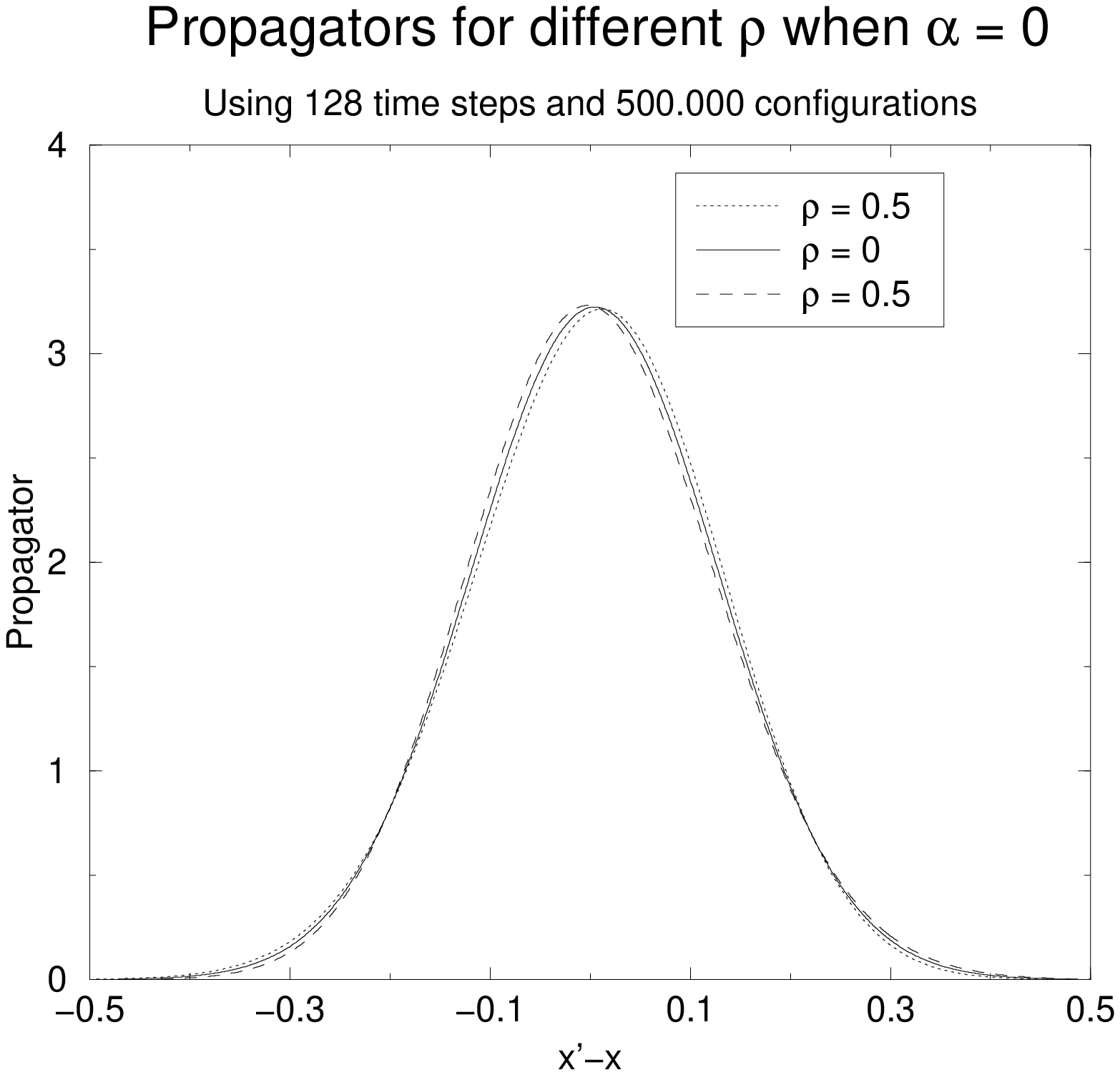, height=7.0cm, width=10cm} \\
\epsfig{file = 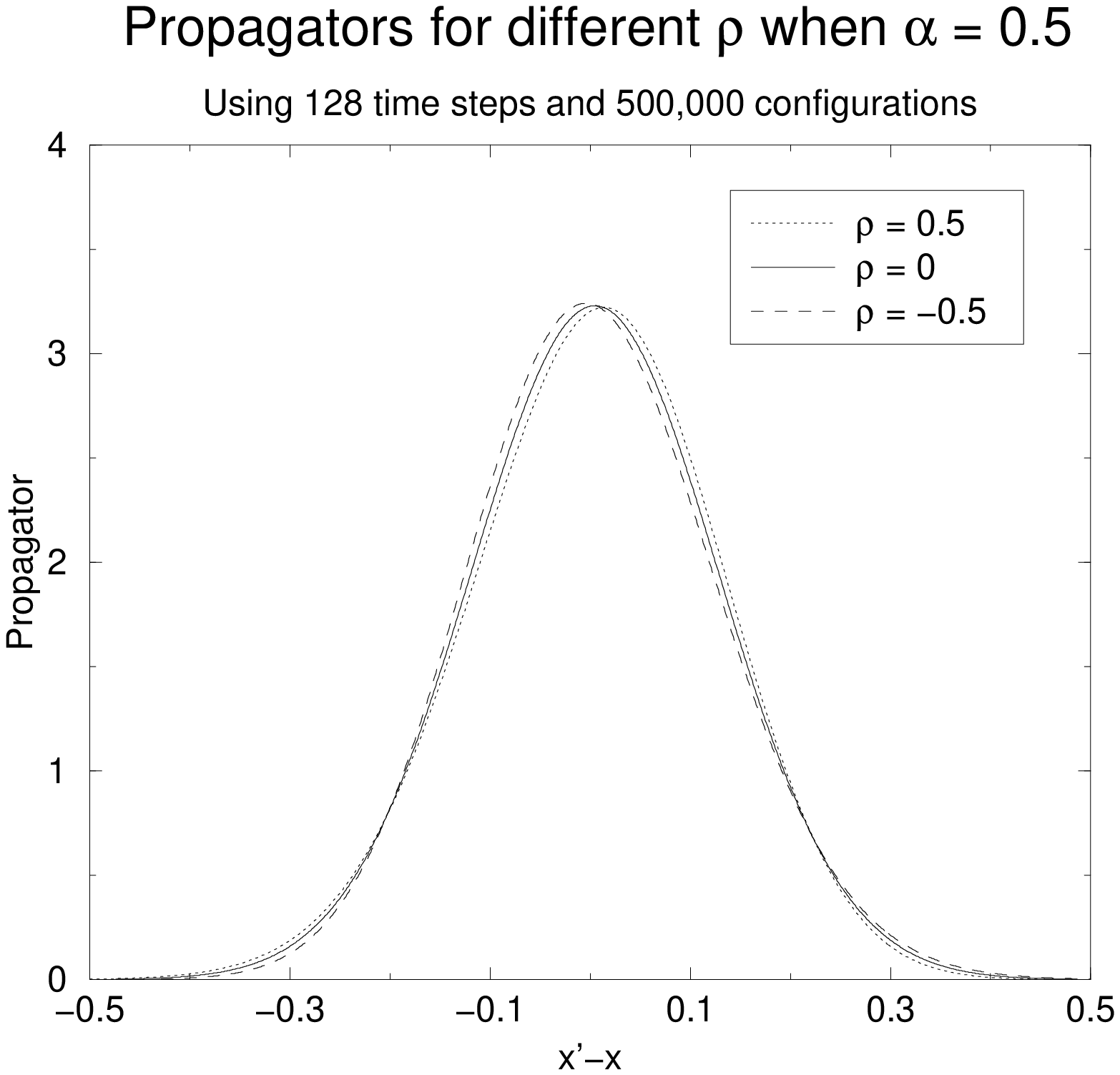, height=7.0cm, width=10cm} \\
\epsfig{file = 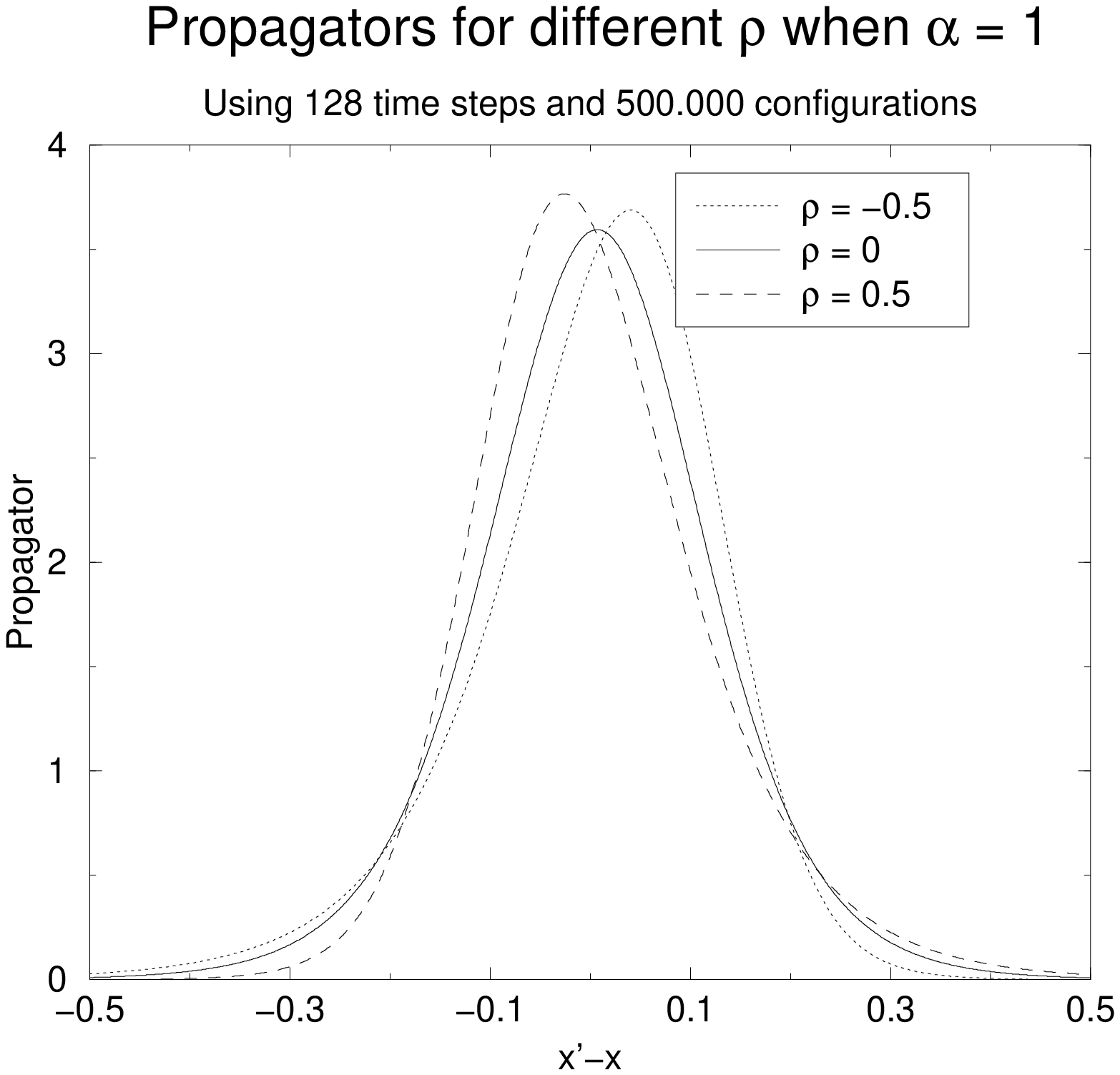, height=7.0cm, width=10cm} \\
\end{tabular}
\end{center}
%\centerline{prop.eps, prop5.eps, prop1.eps}
\caption{Propagators for different $\rho$ when $\alpha = 0,\, 0.5
\text{ and }1$.} \label{Fig_rho}
\end{figure}

This can be explained in terms of the propagator obtained by the
Monte Carlo simulation. From figure \ref{Fig_rho}, we see that the
propagator for positive correlation is greater for very large $x'-
x$ (where $x = \ln S_0$ and $x' = \ln S$) as compared to the
propagator for zero correlation. This essentially means that the
probability for the underlying security price reaching very large
values is higher when the correlation is positive. Hence, the
price of options with a very large strike price is larger when the
correlation between the processes for the underlying security
price and the volatility process is positive. For negative
correlation, we see that the propagator for small, positive $x'-x$
is greater than that for zero or positive correlation. Hence, for
options whose strike price is smaller, there are two competing
factors, the propagator for $x'$ slightly larger than $x$ and the
propagator when $x'>>x$. For relatively large strike prices, the
latter is more important and the implied volatility is higher when
the correlation is positive while for relatively small strike
prices, the former is more important and the price when the
correlation is positive is lesser than for zero or negative
correlation. The same analysis can be performed for the case of
negative correlation and is consistent with the simulated results.

We can also intuitively examine why the propagator has this form
when the two processes are correlated. If the two processes are
positively correlated, we have two possibilities. If the price
initially increases, so will the volatility and hence, large price
increases become more likely while small price increases become
less likely. On the other hand, if the price initially decreases,
so will the volatility and the price is more likely to remain at
that value. Hence, for positive correlation, the propagator must
be higher for prices slightly lower and much greater than the
initial price and lower for prices slightly larger than the
initial price. The reverse holds true for negative correlation.
This naive reasoning is fully borne out in figure \ref{Fig_rho}.

We were unable to simulate frowns or any sort of kinks in the
implied volatility curve for any values of the parameters. This
seems to suggest that stochastic volatility even with arbitrary
correlation places some constraints on the shape of the implied
volatility curve. We can use this to check whether the hypothesis
that the volatility is stochastic is reasonable or otherwise. We
note that the empirical implied volatility curve used for our
calibration does satisfy this criterion.

\clearpage
\clearpage

\subsection{The Effect of Mean Reversion on the Option Price}
Several authors including Heston\cite{heston} and Hull and
White\cite{hullone} have considered mean-reverting processes as
there is some empirical evidence that the volatility follows a
mean-reverting process. We note that our process also includes
mean reversion since $\lambda$ and $\mu$ can be adjusted so that
the volatility performs a mean-reverted process. We find that the
effect of mean reversion is straightforward in that it only seems
to change the implied volatility curve so that it moves closer to
the mean value.

\begin{figure}[ht]
\begin{center}
\epsfig{file = 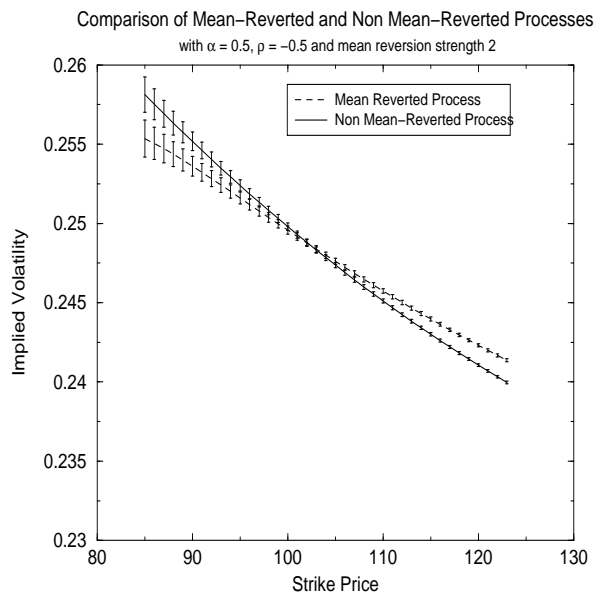, height=8cm, width=8cm}
\end{center}
%\centerline{mean.eps}
\caption{Comparison of a mean-reverting process and a non
mean-reverting process} \label{Mean_Revert}
\end{figure}

For example, consider figure \ref{Mean_Revert}. In this case, we
used an initial volatility of $V_0 = 0.0625$ (so that $\sigma_0 =
0.25$). We set $\lambda = 0.125 = 2 \times 0.0625$ and $\mu = -2$
so that the volatility is performing a mean-reverting process with
the mean the same as the initial value. We see that we indeed
obtain the expected behaviour as compared to the non
mean-reverting process. The mean reverting process just produces
an implied volatility curve of a similar shape which is  closer to
the mean value.

\begin{figure}[htbp]
\begin{center}
\epsfig{file = 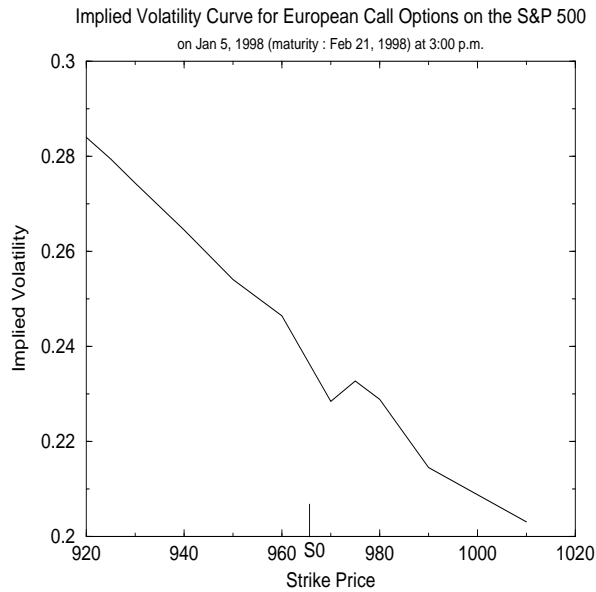, height = 8cm, width = 8cm}
\end{center}
%\centerline{market.eps}
\caption{The implied volatility curve for European call options on
the S\&P 500 Index maturing on Feb. 21, 1998 on Jan. 5, 1998 at 3
p.m.} \label{Fig_market}
\end{figure}

\begin{table}[ht]
\begin{center}
\begin{tabular}{|c|c|} \hline
Strike Price & Option Price\\ \hline 920 & 68.0\\ 925 & 64.125\\
930 & 60.25\\ 940 & 52.75\\ 950 & 45.5 \\ 960 & 39.0\\ 970 & 31.5
\\ 975 & 29.75\\ 980 & 27.0\\ 990 & 21.0
\\ 1010 & 13.0 \\ \hline
\end{tabular}
\end{center}
\caption{The prices of European call options on the S\&P 500 Index
whose maturity was on 21 Feb, 1998 on Jan 5, 1998 at 3:00 p.m. The
prices were taken to be those of the closest trade if there was a
trade within half an hour and the average of the bid and ask
prices otherwise.} \label{Market_data}
\end{table}

\subsection{Calibration with Market Data} We compare our model with
market data to see how well it works. Since volatility information
is available only as 10 day or 50 day averages and the options we
were comparing the market data to had 47 days to expiration, we
used the initial volatility as a free parameter. The market data
used were the prices of European call option on the Standard and
Poor's 500 Index at 3 p.m. on Jan 5, 1998 with the maturity date
given as Feb 21, 1998. The prices taken were either the trade
nearest to 3 p.m. if it was within half an hour and the average of
the bid and ask prices closest to 3 p.m. otherwise. The data are
presented in table \ref{Market_data}. We note that the number of
option prices we have is much larger than the number of free
parameters in the model. The value of the Standard and Poor's 500
Index at the same time was given as 965.61 (since the S\&P 500 is
traded several times every minute on the average, obtaining data
for it presented no problem). The risk-free interest rate $r$ was
5.131\% and the annualized dividend yield was 1.617\% at the same
time. The implied volatility curve for the market data is shown in
figure \ref{Fig_market}.

Looking at the market data, we immediately see that the implied
volatility is almost monotonically declining. Hence, according to
our model, the correlation is very probably negative. We also see
that the implied volatilities vary within quite a wide range
implying that $\xi$ must be quite high as the volatility must vary
widely for the implied volatility to do so. Further, the value of
the initial volatility can be seen to be about 25\% (according to
our model, not the actual initial volatility which we could not
determine with reasonable accuracy).

\begin{figure}[ht]
\begin{center}
\epsfig{file = 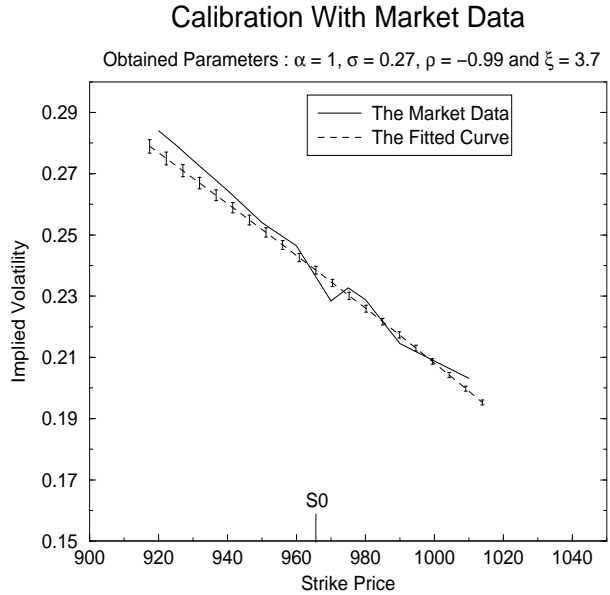, height = 8cm, width = 8cm}
\end{center}
%\centerline{compare.eps}
\caption{The above graph shows the implied volatility curve
produced by the fitted values together with the market data.}
\label{Calibrat}
\end{figure}

Since there is no simple functional form for the option price, the
calibration was performed manually. The reasoning above enabled us
to start with fairly accurate values. Thus, while we cannot
guarantee that the result is the best fit curve in any precise
sense, we can see from figure \ref{Calibrat} that the fit is very
good. While the in the money options seem to not fit so well, this
might be because these options are thinly traded.

The skeptical reader might comment that the number of free
parameters (4) is quite large and that a fairly wide range of
empirical curves might be fitted. However, we note that our model
predicts only three possibilities for the implied volatility
curve, namely a ``smile'' (low correlation), monotonically
increasing (positive correlation) or monotonically
decreasing\footnote{The simulations for non-zero correlation show
more interesting behaviour but we are interested only in the
behaviour for strike prices close to the underlying security price
as these are the only kind of options traded in the market}
(negative correlation). The existence of a ``frown'' or kinks in
the implied volatility curve would be disastrous for the model.
(There does appear to be a small kink in the empirical curve but
the scale of the kink is very small and occurs for only one
value.)

Indeed, as shown in figure \ref{Calibrat}, some calibrated values
of the parameters for European call otpions on the S\&P 500 Index
on 21 Feb. 1998 are $\sigma = 0.27$, $\rho= -0.99$, $\xi = 3.7$
and $\alpha =1$.

%%%%%%%%%%%%%%%%%%%%%%%%%%%%%%%%%%%%%%%%%%%%%%

\section{Numerical Algorithm for $\alpha =1$}\label{num}

In our previous simulation, we found that the value of $\alpha$
for many typical market data is approximately one, that is $\alpha
\approx 1$. It turns out that for this case we have very efficient
algorithms, namely the bisection method.  This algorithm is of
intrinsic interest and allows us to examine the behavior of
stochastic volatility in great detail.  In particular, we
investigate the effect of $\rho$ on implied volatility and show
how smiles turn to frowns and so forth.  We have intentionally
chosen the maturity time to be one year so as to magnify the
effects. To consider the numerical simulation of our model for
$\alpha = 1$ and $\lambda=0$, we first set $\alpha = 1$ in
eq(\ref{propagator}) to eq(\ref{szero}) to get
\begin{equation}
<x,y| \expon^{- \tau H}| x^\prime> = \int DY \dis
\frac{\expon^{S_0 +
S_1}}{\sqrt{2\pi\epsilon(1-\rho^2)\sum_{i=1}^N\expon^{y_i} }}
\label{func} \end{equation} where %%
\bsq S_0 & = & -\frac{\epsilon}{2\xi^2} \sum_{i=1}^{N}
(\frac{\delta y_i}{\epsilon} + \mu -
\frac{\xi^2}{2})^2 \\ %%
S_1 & = & -\dis \frac{1}{2(1-\rho^2) \epsilon \sum_{i=1}^{N}
\expon^{y_i}} \left\{ x - x^\prime + \epsilon \sum_{i=1}^{N} (r -
\frac{1}{2} \expon^{y_i} ) \right.
\nonumber\\ %%
& & \mbox{\hspace{15mm}} \left. -\frac{\rho}{\xi} \sum_{i=1}^{N}
\expon^{\frac{y_i}{2}} \left[ \delta y_i +
\epsilon(\mu - \frac{\xi^2}{2})\right] \right\}^2 \\%%
 \delta y_i &
= & y_i - y_{i-1}, \mbox{\hspace{1cm}} y_N = y = \mbox{{\rm
initial volatility.}} \label{ess} \esq %%
In eq(\ref{func}), we have used the notation $\int DY$ to mean
product $dy_0 \left(\prod_{i=1}^{N-1} \int_{-\infty}^{\infty} \dis
\frac{dy_i}{\sqrt{2\pi\epsilon\xi^2}}\right)$.

Moreover, the complete information regarding the dynamics and
evolution of stock price $S(t)$ and its volatility $V(t)$, their
cross correlators as well as the fluctuations is given by the
discrete time path integral equation \bea <x,y| \expon^{- \tau H}|
x^\prime> & = & \int_{-\infty}^\infty dy^\prime p(x,y, \tau
|x^\prime, y^\prime) \nonumber \\ & = & \lim_{N \rightarrow
\infty} \int DX DY \expon^{S^\prime} \label{eqn19} \eea where, for
$\epsilon = \frac{\tau}{N}$, $$ \int DX  =  \dis
\frac{\expon^{-\frac{y_N}{2}}}{\sqrt{2 \pi \epsilon (1 - \rho^2)}}
\prod_{i = 1}^{N - 1} \int_{-\infty}^{\infty} \frac{dx_i
\expon^{-\frac{y_i}{2}}}{\sqrt{2 \pi \epsilon (1 - \rho^2)}} $$
and where the `action' $S^{\prime}$ is defined by \bea S^\prime &
= & - \frac{1}{2} \epsilon \sum_{i = 1}^{N} \left\{ \frac{1}{
\xi^2} (\frac{\delta y_i}{\epsilon} + \mu - \frac{1}{2} \xi^2)^2
\right. \nonumber \\ &  & \left. + \frac{\expon^{-y_i}}{(1 -
\rho^2)} \left[ \frac{\delta x_i}{\epsilon} + r - \frac{1}{2}
\expon^{y_i} - \frac{\rho}{\xi} \expon^{\frac{y_i}{2}}
(\frac{\delta y_i}{\epsilon} + \mu - \frac{1}{2} \xi^2) \right]^2
+ O(\epsilon) \right\} \label{ess2} \eea

To determine numerically the probability $P(x, y, \tau| x^\prime,
y^\prime)$, one could use the Metropolis method \cite{bhanot} to
evaluate numerically the path integral given in eq(\ref{func}) by
finding the expectation value of $\dis
\frac{\expon^{S_1}}{\sqrt{2\pi\epsilon(1-\rho^2)\sum_{i=1}^{N}
\expon^{y_i}}}$, the functional average being performed over
configurations of $y_n$ with a probability distribution given by
$\dis \expon^{S_0}$, where $S_0$ and $S_1$ are given by the
formulae in eq(\ref{ess}a) and eq(\ref{ess}b) respectively.
However, given the special form of $S_0$ which can be interpreted
as the kinetic energy of a free quantum particle, a more efficient
method is to generate configurations based on the bisection method
\cite{roep}. In the bisection method, unlike the Metropolis case,
all configurations generated are accepted and this is therefore a
better algorithm.

In the bisection method, the interval $\tau$ is first divided into
half; the sample value of $y$ at the center is generated by $\dis
y(\frac{\tau}{2}) = \frac{1}{2} \{ y(0) + y(\tau) + \tau^{1/2} z
\},$ where $z$ is a standard normal variate, $N(0,1)$, namely a
normal random variable with mean zero and variance of unity.  The
sub-intervals $[0,\tau/2]$ and $[\tau/2, \tau]$ are further
bisected and the above algorithm is repeated and so forth.  In
general, the interval $[a,b]$ is bisected and the sample value at
the mid-point is given by $\dis y(\frac{a + b}{2}) = \frac{1}{2}
\{ y(a) + y(b) + (b-a)^{1/2} z \}$, where $y(a)$ and $y(b)$ are
the values of $y(t)$ at the points $a$ and $b$ respectively.
After $N$ bisections, we obtain a time lattice with $2^N$ discrete
points with spacing $\epsilon=\frac{\tau}{2^N}$; and the average
of $\dis
\frac{\expon^{S_1}}{\sqrt{2\pi\epsilon(1-\rho^2)\sum_{i=1}^{N}
\expon^{y_i}}}$ is taken over the configurations generated.

\subsection{Pseudo-Codes for the Algorithm}

Essentially, the main formula that we used to simulate the
derivative pricing is eq(\ref{func}).   To compute the derivative
prices, the program in our simulation can be broken into three
major steps:
\begin{itemize}
\item Generate $2^N ~ ~ y_i$ variables using the bisection method,
\item Evaluate the path integration using eq(\ref{func}),
\item Compute the derivative price.
\end{itemize}

The first step in the computation requires the generation of $2^N$
variables using the bisection method.  The pseudo-code for the
bisection subroutine is as follows: {\small
\begin{itemize}
\item Initialize an array of size $2^N +1$ (Call it Rand Y($i$), for random
$y_i$)
\item Set $y_1$ and $y_M$ where $M=2^N+1$ to the initial
and end values of the volatility.
\item Do J=1 to m
\item Do I=1 to $2^{J-1}$
\item Call Normal Variate $N(0,1)$, Norm
\item Rand Y($\dis \frac{M(2^I-1) + (2^J-2^I+1)}{2^J}$) $\leftarrow$
$\frac{1}{2} \{$ Rand Y($\dis \frac{M(I-1) +
(2^{J-1}-1+1)}{2^{J-1}}$)
\\ \mbox{\hspace{3cm}}
 + Rand Y($\dis \frac{IM+(2^{J-1}-1)}{2^{J-1}}$) +
 $\dis Norm(\sqrt{\frac{\tau}{2^{J-1}}}) \}$
\item End of I Loop.
\item End of J Loop.
\end{itemize}
}

Computation of the path integration using eq(\ref{func}) is
straightforward. We first fix $x$ and $y$ to be some initial value
of stock price and volatility.  We then divide the variable $
x^\prime$ into $N$ divisions\footnote {Typically, due to the
exponential relation of the variables with the stock price, a
range between $\pm 7$ should be sufficient.} and store them as
arrays. For each value of $x^\prime$ and $y^\prime$, one then
calls the subroutine Bisect which essentially provides a
normalized array of $2^M+1$ points and computes using
eq(\ref{ess}b), the average of $\dis
\frac{\expon^{S_1}}{\sqrt{2\pi\epsilon(1-\rho^2)\sum_{i=1}^{N}
\expon^{y_i}}}$. This procedure effectively yields the probability
in eq(\ref{func}). As a final step, one computes the simulated
derivative pricing by integrating over $y^\prime$ using a simple
integration algorithm based on Bode's rule. This integration is
exact for any polynomial up to and including degree 5.

\section{Results of Numerical Simulations for $\alpha=1$ and $\lambda=0$}\label{disc}

\subsection{Parameters}

There are a number of parameters which we can fix freely without
affecting the analysis of the behaviors in the implied volatility
curves. Based on the current market data, we have set the interest
rate arbitrarily at a plausible rate of 6 \% per annum throughout
the simulation. The other free parameters, which are fixed
throughout the simulations, assume the following values:

\begin{tabular}{ll}
Interest Rate, $r$: &  0.06 or 6 \% \\ Time interval, $\tau$: & 1
year
\\ Number of bisections: & 32 \\ Number of points for $x^\prime$
integration: & 100 \\ Initial Volatility, $V$: & 1 (i.e. $ y_N=0$)
\end{tabular}

\subsection{Simulations}
Under the path integral formalism, one can vary the new parameters
like $\rho$, $\xi$ and $\mu$ in the model and study its effect on the
option price. With sufficient simulation, one hopes to identify
precisely the nature of these changes and interpret their importance to
the fluctuation of the option prices in the market.
Like Black-Scholes model, one can also get closed form expressions of the
option prices under certain constraints and limits.  Unlike
Black-Scholes model, it does not assume constant
volatility and thus provides greater
flexibility for a calibration of the market data.

We shall see that the variation of the call option prices against
strike prices for different values of correlation parameter, $\rho$ can
differ significantly from Black-Scholes case.  We simulate option prices
with stochastic volatility using a stock price of 100.
In figure \ref{vc1}, we
have plotted the two graphs of call option prices against strike price
for a fixed $\rho= 0.001$ but differing values of $\xi$.  In this
simulation, we have set $\mu = 0$. In one graph,
we have kept $\xi$ as 0.01 while in the other graph we have fixed the
same parameter as 0.001.

\begin{figure}[htbp]
%\centerline{  \epsfig{ file=vc2a.eps, height=6cm} }
\centerline{graphic1.gif} \caption{Graph of the call option price
with stochastic volatility against strike prices for $\mu =0$ and
$\rho=0.001$ is plotted. A similar graph computed from
Black-Scholes equation is also drawn for comparison. }\label{vc1}
\end{figure}

A graph of the option prices computed from Black-Scholes model
with a variance of unity is also provided. We observe that despite
the small value of $\rho=0.001$, there are still some drastic
differences between the option prices simulated with stochastic
volatility and the standard Black-Scholes equation. The difference
seems drastic for strike prices below the at-the-money position of
100 units but for strike prices above the at-the-money position,
this difference may not be large. We have checked numerically that
the option price with stochastic volatility converges, as
expected,
 to that
predicted by the
Black-Scholes formula in the limit $\xi \rightarrow 0$.

We simulate and calculate the implied volatilities for various
values of $\rho$ holding stock price constant at 100 and then
allowing the strike prices to vary. Our initial simulation are
based on values of $\rho$ between 0.1 and 0.9. We believe that one
can glean important information on the {\it behaviour} of the
curves with these values. The values of $\mu$ and $\xi$ have again
been arbitrarily fixed at 0.1. The simulated data for $\rho=0.1$
to $\rho=0.3$ is tabulated and shown in table \ref{raw1}. The
implied volatility curves for $\rho = 0.1, 0.2, 0.3$ are plotted
against different strike prices in figure \ref{vc2}.  The graphs
clearly show that the volatility curves assume frowns for these
positive values of $\rho$.

\begin{figure}[htbp]
%\centerline{ \epsfig{file=rho.eps, height=8cm} }
\centerline{graphic2.gif}
\caption{Graphs of implied volatility
curves for positive $\rho$ values between $0.1$ and $0.3$. The
parameters $\xi$ and $\mu$ are set to 0.1 and the stock price is
fixed at 100.  These curves appear as `frowns'. }\label{vc2}
\end{figure}

\begin{table}[ht]
\centerline{
\begin{tabular}{||l|l|l|l|l|l|l||}
\hline
\hline
$\mu$ & \multicolumn{2}{|c|}{0.1} &
\multicolumn{2}{|c|}{0} &
\multicolumn{2}{|c|}{0} \\
\hline
 & \multicolumn{2}{|c|}{$\rho = 0.1, \xi=0.1$} &
\multicolumn{2}{|c|}{$\rho = 0.01,  \xi = 0.01$} &
\multicolumn{2}{|c|}{$\rho = -0.01, \xi = 0.01$} \\
\hline
Strike Price & Option & Volatility &  Option & Volatility
&  Option & Volatility \\
\hline
\hline
75&     31.2748&  0.635 &  38.897&  0.3216& 76.615& 2.16\\
80&     30.6564&  0.702 &  38.216&  0.443&  74.989& 2.119\\
85&     30.0967&  0.749 &  37.351&  0.5212& 73.386& 2.08\\
90&     29.5342&  0.79  &  36.584&  0.5804& 71.726& 2.041\\
95&     29.0275&  0.827 &  35.907&   0.6298& 69.893& 1.995\\
100&    28.5038&  0.861 &  35.318&   0.6707& 68.412& 1.967\\
105&    27.9804&  0.888 &  34.648&  0.7057& 66.718& 1.93\\
110&    27.5226&  0.911 &  33.992&   0.7377& 65.213& 1.902\\
115&    27.0136&  0.931 &  33.283&   0.7645& 63.628& 1.871\\
120&    26.5956&  0.95  &  32.666&   0.7906& 62.134& 1.844\\
125&    26.174&   0.971 &  32.217&  0.8138& 60.737& 1.821\\
 \hline
 \hline
\end{tabular}
}
\caption{Simulated data values of call option prices and their implied
volatilities for different strike prices with varying $\rho$ values and
for a stock price of 100 }\label{raw1}
\end{table}

The parameter $\rho$ measures the amount of correlation between
the stock price and its volatility. Now, correlation coefficient
can assume negative values and we should not disregard this
possibility. Also, negative values of $\rho$ tends to `whip' up
the value of the integrand in eq(\ref{ess}) or eq(\ref{ess2}). We
have seen from figure \ref{vc2} that the volatility curves for
positive $\rho$ concave downwards as `frowns'. At this stage, we
may be tempted to surmise that concavity and convexity of
volatility curves are associated with positive and negative values
of $\rho$ respectively. We shall soon see that this is not true.

We investigate the volatility curves for negative $\rho$. For the
case in which the stock price is fixed at 100 and the parameters
$\mu=\xi$ are set to 0.1, the graphs of the implied volatilities
against strike prices for $\rho= -0.1, -0.2, -0.3$ are given in
figure \ref{pnrho}. The volatility graph for $\rho=-0.1$ certainly
concave upwards as `smiles'.  However, the graphs for $\rho=-0.2$
and $\rho=-0.3$ flip over and appear as frowns again. Thus,
negative values of $\rho$ do not necessarily yield volatility
curves which concave upwards as smiles.

\begin{figure}[htbp]
%\centerline{  \epsfig{file =vlnrho.eps, height=8cm} }
\centerline{graphic3.gif} \caption{Implied volatility curves for
negative $\rho$ values. The parameters $\mu$ and $\xi$ are fixed
at 0.1.} \label{pnrho}
\end{figure}

Under path integral formalism, it is always possible to expand the
probability given in eq(\ref{func}) perturbatively in powers of
$\xi$ when the values of $\xi$ are in general very small
\footnote{Strictly speaking, the values of the strike prices and
stock prices must also be appropriately tuned before a
perturbative approach is possible.} and evaluate exact closed-form
expression for the coefficients in the expansion \cite{hull2}.
However, when $\xi$ assumes typically the order of unity, ($\xi
\sim 1$ which is what market data indicates) it is generally
theoretically impossible to perform such an expansion and the
perturbative method fails.

However, unlike perturbative analysis, the path integral formalism
still permits numerical evaluation for these values of $\xi$. Some
simulated results for large $\xi=1$ with $\mu = 0.1 $ are shown in
figure \ref{perb}. We also investigate the variation of the
implied volatility curves against strike prices at different
initial stock prices.  When we fix the values of $\mu$ and $\xi$
to 0.1. with $\rho= -0.1$, the volatility curves concave upwards
as smiles for each stock price. In figure \ref{stkp}, we show our
results for four different values of stock prices, namely 50, 75,
105 and 200.  Note that the volatility curves for the last two
stock prices, namely 105 and 150, appear to coincide with each
other. We also note that the large value of $\mu$ which we have
chosen for the model may have unnecessarily distorted the graphs
in figure \ref{stkp} and shifted them away from the at-the-money
position.

\begin{figure}[htbp]
%\centerline{ \epsfig{file =largemu.eps, height=8cm} }
\centerline{graphic4.gif} \caption{Simulation of Implied
Volatility curves for large values of $\xi ( \xi = 1)$ in which
perturbation analysis fails. The graphs in this figure are
simulated $\mu=0.1$ with varying $\rho$ values. } \label{perb}
\end{figure}

\begin{figure}[htbp]
%\centerline{  \epsfig{file=stkp.eps,height=8cm} }
\centerline{graphic5.gif} \caption{Graphs of implied volatility
against strike price for varying values of stock price, $S$. In
this simulation, the parameters $\mu$, $\xi$ and $\rho $ have been
arbitrarily fixed at 0.1, 0.1, -0.1} \label{stkp}
\end{figure}

In most mature market, the value of $\rho$ can be small for some
securities. We simulate the volatility curves with $\mu$ set to
zero and with the parameters $\rho$ and $\xi$ held small at -0.02
and 0.01 respectively. In this case, we note two interesting
observations. Firstly, negative $\rho$ values do not necessary
lead to volatility curves which concave upwards as smiles. As seen
in figure \ref{blk}, the concavity of the volatility curves can
vary with different stock prices. Secondly, if we plot the option
prices generated in the simulation with a stock price of 100 and
compare the graph with Black-Scholes model as in figure \ref{bsm},
we observe that the two graphs intersect near the strike price of
100.  This result is consistent with the observations in some
models \cite{stein} that stochastic volatility appear significant
away from the at-the-money position.  This means that implied
volatility is lowest at-the-money and increases as the strike
prices moves away from that position.

\begin{figure}[htbp]
%\centerline{\epsfig{file=black.eps,height=8cm} }
\centerline{graphic6.gif} \caption{Graph of option price with
$\mu$, $\xi$ and $\rho $ set at 0, 0.01, -0.02 and the stock price
fixed at 100 compared to the Black-Scholes model with the same
initial volatility. We have also indicated the at-the-money
position using a vertical line at the strike price of 100.}
\label{bsm}
\end{figure}

The initial volatility in our simulation has been fixed at 1 so that we
can see the full effects of volatility to the option prices.  Generally
we would anticipate a smaller value of initial volatility for the
market data.
We have also performed simulations using
a smaller initial volatility of 0.2.
Figure
\ref{compslv} shows the different implied volatility curves which we
have obtained using this smaller initial volatility for $\rho=0.01$ and
$\xi=-0.01$ with $\mu=0$ at two different stock prices, namely
$S=50(l)$ and $S=150(l)$.  On the same graph, we have also shown the
implied volatility curves for the two stock prices, $S= 50(h)$ and $S=
150(h)$ for an initial volatility of unity.  We note that the curves can
differ significantly for different values of initial volatility. Indeed,
when we increase the stock price from  50 to 150, we
note that the implied volatilities generally increase for different
strike price at the lower initial volatility.  However, the reverse
effect seems to be taking place when the stock price increase over the
same range for a high initial volatility.

\begin{figure}[htbp]
%\centerline{\epsfig{file=stkg.eps,height=8cm} }
\centerline{graphic7.gif} \caption{Volatility curves for
different stock price, $S$. In this simulation, the parameters
$\mu$, $\xi$ and $\rho $ have been arbitrarily fixed at 0, 0.01,
-0.02.} \label{blk}
\end{figure}

\begin{figure}[htbp]
%\centerline{  \epsfig{file=compr1.eps, height=8cm} }
\centerline{graphic8.gif} \caption{Volatility curves for two
stock prices, $S=50$ and $S=150$ at different initial
volatilities. For the lower initial volatility of 0.2, we have
denoted the curves by $S=50(l)$ and $S=150(l)$, whereas for the
larger initial volatility of 1, we have labeled the curves
$S=50(h)$ and $S=150(h)$.} \label{compslv}
\end{figure}

Since the results can differ drastically for different initial
volatilities, we have also done some simulation of the implied
volatility curves at an initial volatility of 0.2 for different stock
prices.  Figure \ref{lp1} and \ref{ln3} shows the results of our
simulation of the implied volatility curves for $\xi$ and $\mu$ fixed at
0.01 and 0 respectively.  In figure \ref{lp1}, we have considered a
positive value of $\rho=0.01$ whereas in figure \ref{ln3}, we perform the
simulation with a negative value of $\rho=-0.03$.
\begin{figure}[htbp]
%\centerline{ \epsfig{file=loivn3.eps, height=8cm} }
\centerline{graphic9.gif} \caption{Volatility curves for
different stock prices between $S=50$ and $S=150$ with the initial
volatility set at 0.2. The parameters $\xi$, $\mu$ and $\rho$ have
been set at $0.01, 0$ and $-0.03$ respectively.} \label{ln3}
\end{figure}

\begin{figure}[htbp]
%\centerline{ \epsfig{file=linvp1.eps, height=8cm} }
\centerline{graphic10.gif} \caption{Volatility curves for
different stock prices between $S=50$ and $S=150$ with the initial
volatility set at 0.2. The parameters $\xi$, $\mu$ and $\rho$ have
been set at $0.01, 0$ and $0.01$ respectively.} \label{lp1}
\end{figure}

The concavity of the implied volatility curves can significantly
affect the decisions of market analysts.  In figure \ref{blk} and
\ref{lp1}, one quickly observes the drastic change in the
concavity as the stock prices vary from 50 to 150.  In particular,
it is interesting to investigate how the implied volatility curves
change their concavity between the stock price of 60 and 75 in
figure \ref{blk}. Indeed, figure \ref{blk} shows how a smile can
turn into a frown if the initial stock price falls from 200 to say
50.

\begin{figure}[htbp]
%\centerline{\epsfig{file=fine.eps,height=8cm} }
\centerline{graphic11.gif}
\caption{Volatility curves for
different initial stock prices between $S=63.1$ and $S=64.8$ with
the parameters fixed at the values for the graph in figure 9. The
parameters $\xi$, $\mu$ and $\rho$ have been set at $0.01, 0$ and
$-0.02$ respectively.} \label{fine}
\end{figure}

\begin{comment}
Numerical simulations are only useful if they can duplicate
realistic data from the industry.  As a further numerical check
for our simulation, we have monitored S \& P 500 derivative prices
in the market and observed almost similar behaviors in some of the
volatility curves. As an illustration, we show in figure
\ref{actual} the market data for S \& P 500 call prices using the
ask price for  two days, namely May 27 and May 28 1997. The
prevailing interest rate is 5.098 \%.  The dividend yield for the
index is 1.742 \%.  This gives us an effective interest rate of
3.356 \%. Consequently, the implied volatility curves are computed
with an interest rate of 3.356 \% over a period of 90 days. We
observe that there is strong similarity in the actual implied
volatility curves with some of the implied volatility curves
generated in our data. It is probably interesting to compare the
implied volatility curves in figure \ref{actual} with those
obtained in figures \ref{lp1} and \ref{ln3}.
\end{comment}

\section{Efficiency of Algorithms}

The normal way of doing Monte-Carlo simulations to find option
prices with stochastic volatility involves directly simulating the
process
\begin{align}
dS = rS dt + \sqrt{V} S Wdt\\ dV = (\lambda + \mu V)dt + \xi
V^\alpha Qdt
\end{align}
by discretising it using the Euler method to
\begin{align}
\Delta V_i = (\lambda + \mu V_i)\Delta t + \xi V_i^\alpha
\epsilon_1 \sqrt{\Delta t}\\ \Delta S_i = rS_i \Delta t +
\sqrt{V_i} S_i \epsilon_2 \sqrt{\Delta t}
\end{align}
where the standard normally distributed random variables
$\epsilon_1$ and $\epsilon_2$ with correlation $\rho$ are
generated by first generating two uncorrelated standard normally
distributed random variables $\delta_1$ and $\delta_2$ and using
$\epsilon_1 = \delta_1$ and $\epsilon_2 = \rho \delta_1 +
\sqrt{1-\rho^2} \delta_2$. The final values of $S$ are stored and
the option price with strike price $K$ is estimated by considering
$E[\max(S-K, 0)]$. This is the algorithm used along with the
control variate method in Johnson and Shanno\cite{johnson} (we
henceforth call the above algorithm the standard algorithm).

Before we compare the efficiency of our algorithm with the
standard one, we look at the possible sources of error and how the
error due to each source scales with run time for both. The major
source of error for both algorithms is the Monte Carlo error which
goes as $N^{-1/2}$ and which goes as the square root of the run
time. Another common source of error in both the algorithms is the
discretization of the process for $y$ or $V$. The error in $y$ is
then of the order of $h^{1/2}$ where $h$ is the time step used.
However, the effect of this error on the option price is virtually
impossible to estimate and we verify that the number of time steps
is adequate empirically by comparing two simulations with
different numbers of time steps. The standard algorithm has a
further error due to the discretization of the $S$ process. The
error is again of the order of $h^{1/2}$. The effect of this error
on the option price is of the same order as the payoff of the
option is piecewise linear with respect to $S$. Hence, this error
also goes inversely as the square root of the run time. Our
algorithm has other errors due to the finiteness of the
integration over $x$, the interpolation error and the quadrature
error which goes as $h^4$. The error due to the finiteness of the
quadrature domain can be made completely negligible with minimal
effort as the propagator goes as $e^{-d^2}$ where $d = x-x'$.
Hence, for large enough $d$, this error goes as $e^{-t^2}$ where
$t$ is the run time. This is a truly negligible error. The
interpolation error is difficult to estimate but we empirically
show that it is very small for interpolation over 100 points as
the propagator is very smooth. The quadrature error goes as $h^4$
(Simpson's rule). Since most of the computer time is spent on the
Monte Carlo simulation, we can also make this error very small
with only a small increase in run time. Hence, we see that one of
the main advantages of our algorithm is that it has a
significantly lower error for the same number of configurations
(since it has no error due to the $S$ simulation since these
degrees of freedom have been integrated out).

When generating a single set of option prices with the same error
tolerance, our algorithm is about 30 times faster. However, our
algorithm has an important advantage in that $t_1$ and $t_2$ are
independent of $\rho$. Hence, when we generate sets of data with
all parameters except $\rho$ fixed, we only need to calculate the
new propagator using the terms and integrate over the final
payoff. This effectively results in an increase in efficiency of a
factor of six to seven when we calculate the prices for 10
different values of $\rho$. Hence, when generating data with
several values of $\rho$, our algorithm is about two hundred times
faster.

At $\alpha =1$, we can use an alternative algorithm called the
bisection method. Compared to usual Monte Carlo methods, the
bisection method provides a reasonably fast algorithm for the
simulation of derivative pricings.   We need to integrate the path
integral over $x^\prime$ values. We first divide the interval for
the $x^\prime$ values into 100 divisions.  For each value of
$x^\prime$, we apply the bisection method and divide the interval
$y$ into $2^5 = 32$ points.  This application of the bisection
method yields one configuration of points. In the algorithm, we
sample $N$  such configurations for each value of $x^\prime$.
Using Bode's rule as an integration algorithm, we finally
integrate over the $x^\prime$ values.  In general, we find that it
is sufficient to execute only $N = 100$ configurations for each
value of $x^\prime$ in the program since the gain in numerical
precision is not substantial. These results were computed with 100
values of strike prices.

\section{Conclusion}

The path integral formalism provides an additional tool for
analyzing derivative pricing. Although our model may seem
heuristic\footnote{As pointed out by one referee, the issue of the
financial market completeness is rather subtle. Allowing trading
in stock and riskless bonds implies incompleteness in the market.
}, our techniques and computational method is generally quite
novel in the financial market and it has several added advantages.
Firstly, it does not assume constant volatility and should be a
better tool for analyzing market data. Secondly, the additional
parameters clearly allow us to study more carefully the behavior
of the volatilities in the market.  And lastly, the formalism
provides a reasonably fast algorithm for computing the option
prices.

In our simulation, we relied on market data to extract the
plausible values of $\mu$, $\alpha$, $\rho$ and $\xi$. Although
the value of $\rho$ in most of the currently well-established
markets is believed to hover between -0.5 and 0.5, we did not
restrict our $\rho$ to these values.  However, it is possible for
$\rho$ to assume higher values in more mature markets. Indeed, we
have treated our simulation largely as an experimental tool to
investigate the qualitative behavior for the implied volatility.
We believe that this new formalism, with proper calibration, can
offer a formidable tool for getting a more accurate pricing of
options. Further, we studied the time period from 90 days to one
year and set the initial volatility arbitrarily at unity.

One of the most important qualitative result from our simulations
for $\alpha \neq 1$ and $\alpha=1$ is that at-the-money, the
implied volatility is equal to the naive fixed volatility of
Black-Scholes, and the latter is approximately equal to the
historical volatility.  Deep-in-the-money and
deep-out-of-the-money show frowns and smiles for implied
volatility depends very much on the value of $\rho$. Finally, the
sharp cross-over for the implied volatility for certain special
values of initial stock price shows that hedging the option for
these values could lead to large changes in the value of the
portfolio and can consequently leads to greater risk.

Finally, we emphasize that it is very important to calibrate this
path integration model with the market data for different
securities and extract all the relevant information about the
range of parameters. Such an approach can be extremely interesting
and revealing for the traders who would like to have some
qualitative ideas regarding the actual behavior of derivative
pricing in the markets.

%\section{Acknowledgement}
\section{Endnotes}
We are grateful to Lawrence Ma (Man-Drapeau, Singapore) and
Choon-Peng Toh (Bank of America) for discussions and helpful
conversations. We also thank G. Bhanot for their helpful comments
and advices regarding the computational aspect of our work.

\appendix
\newpage
\section*{Appendix}
\section{A Quantum Mechanical Formulation of the Problem} We
define the Hamiltonian operator as
\begin{equation}
\begin{split}
\hat{H}(x, y) =& - \left(r-\frac{e^y}{2}\right)\pdif{}{x} -
\left(\lambda e^{-y} + \mu -
\frac{\xi^2}{2}e^{2y(\alpha-1)}\right)\pdif{}{y} -
\frac{e^y}{2}\pdiftwo{}{x} - \rho \xi e^{y(\alpha -
1/2)}\frac{\partial^2}{\partial x \partial y} \\ &- \frac{\xi^2
e^{2y(\alpha - 1)}}{2}\pdiftwo{}{y}.
\end{split}
\end{equation}
The Hamiltonian is non-Hermitian and accounts for the
irreversibility of the stochastic processes in finance. From
eq(\ref{mgeq}), we obtain the Merton-Garman-Schr\"odinger equation
\begin{eqnarray}
\pdif{f}{t} & = &  (r + \hat{H}(x, y))f, \\
 f(x, y, T) & = & \max(e^x-K,0)
\end{eqnarray}
which can be formally solved as
\begin{equation}
\begin{split}
f(x, y, t) &= e^{-r\tau}\int_{-\infty}^\infty dx'  \matel{x,
y}{e^{-\hat{H}\tau}}{x'}f(x', T)\\ f(x, T) &= \max(e^x-K, 0),\,
\tau = T-t
\end{split}
\end{equation}

While this looks deceptively simple, no analytic solution has been
obtained for this equation. The special case $\alpha =
\frac{1}{2}$ was solved using a series method by Hull and
White\cite{hull3} and using elementary probability techniques by
Heston\cite{heston}. the case $\alpha = 1$ was solved by Baaquie
\cite{baaq}.

\section{The Lagrangian for the Problem} The central quantity
whose knowledge is sufficient to solve the problem is the
conditional probability given by the propagator
\begin{equation}
P(x, y, T \mid x', t) = \matel{x, y}{e^{-\hat{H}\tau}}{x'}
\end{equation}
which can be conveniently handled in the Lagrangian formulation of
quantum mechanics.

To determine a Lagrangian for the problem, we discretize time so
that there are $N$ time steps.The time step is then $\epsilon =
\frac{\tau}{N}$. The continuous variables $x(t)$ and $y(t)$ are
discretized to $x_i$ and $y_i$ where $0 \le i \le N$. The operator
$\matel{x, y}{e^{-\hat{H}\tau}}{x', y'}$ can then be decomposed to
\begin{equation}
\begin{split}
\matel{x, y}{e^{-\hat{H}\tau}}{x'} = &\int_{-\infty}^\infty
dy'\matel{x, y}{(e^{-\hat{H}\epsilon})^N}{x',y^\prime} \\ =
&\int_{-\infty}^{\infty} dx_{N-1} \int_{-\infty}^\infty dy_{N-1}
\dots \int_{-\infty}^\infty dx_1 \int_{-\infty}^\infty dy_1
\int_{-\infty}^\infty dy_0 \times\\ &\matel{x_N,
y_N}{e^{-\hat{H}\epsilon}}{x_{N-1}, y_{N-1}} \dots \matel{x_1,
y_1}{e^{-\hat{H}\epsilon}}{x_0, y_0}
\end{split}
\end{equation}
where $x_N = x$, $y_N = y$, $x_0 = x'$ and $y_0 = y'$.

We see that if we can find $\matel{x, y}{e^{-\hat{H}\epsilon}}{x',
y'}$, we can find the propagator and hence the option price.
Therefore, let us look at this quantity more closely. Before we
consider this quantity for the stochastic volatility case, let us
consider the Black-Scholes (constant volatility) case as it is
simpler and retains the essential features.

In the Black-Scholes case, we only have one variable $x$ (as $y$
is just a constant). We write
\begin{equation}
\matel{x}{e^{-\hat{H}_{BS}\epsilon}}{x'} =
N_{BS}(\epsilon)e^{\hat{L}_{BS}\epsilon}
\end{equation}
where $N(\epsilon)$ is a normalization constant. We see that
\begin{align}
&N_{BS}(\epsilon) = \frac{1}{\sigma \sqrt{2\pi \epsilon}}\\
&\hat{L}_{BS} = -\frac{1}{2\sigma^2}\left(\frac{\delta
x}{\epsilon} + r - \frac{\sigma^2}{2}\right)
\end{align}
where $\delta x = x-x'$.

For the stochastic volatility case, we have
\begin{equation}
\begin{split}
N(\epsilon)e^{\hat{L}\epsilon} &= \matel{x,
y}{e^{-\hat{H}\epsilon}}{x', y'}\\ & = \int_{-\infty}^{\infty}
\frac{dp_x}{2\pi} \int_{-\infty}^{\infty} \frac{dp_y}{2\pi}
\matel{x, y}{e^{-\hat{H}\epsilon}}{p_x, p_y} \innprod{p_x,
p_y}{x', y'}
\end{split}
\end{equation}
The Hamiltonian in the phase space basis is given by
\begin{equation}
\begin{split}
\hat{H} =& \frac{e^y}{2}p_x^2 + \xi \rho e^{y(\alpha - 1/2)}p_xp_y
+ \frac{\xi^2 e^{2y(\alpha - 1)}}{2}p_y^2\\ &+ \left(\frac{e^y}{2}
- r - \frac{\delta x}{\epsilon}\right)ip_x +
\left(\frac{\xi^2e^{2y(\alpha-1)}}{2} - \lambda e^{-y} - \mu
-\frac{\delta y}{\epsilon}\right)ip_y
\end{split}
\end{equation}
Hence, we have
\begin{equation}
\begin{split}
N(\epsilon) e^{\hat{L}\epsilon} =& \int_{-\infty}^\infty
\frac{dp_x}{2\pi} \int_{-\infty}^\infty \frac{dp_y}{2\pi}
\exp\left(-\epsilon \left(\frac{e^y}{2}p_x^2 + \xi \rho
e^{y(\alpha - 1/2)}p_xp_y + \frac{\xi^2 e^{2y(\alpha -
1)}}{2}p_y^2\right.\right. \\ &\left. \left. - \xexpr ip_x -
\yexpr ip_y \right) \right).
\end{split}
\end{equation}
We obtain in a straightforward but tedious manner
\begin{equation}
N(\epsilon) = \frac{e^{y(1/2-\alpha)}}{2\pi\epsilon\xi
\sqrt{1-\rho^2}}
\end{equation}
and
\begin{equation}
\begin{split}
\hat{L} = &-\frac{e^{2y(1-\alpha)}}{2\xi^2(1-\rho^2)} \yexpr^2\\ &
+ \frac{\rho e^{y(1/2-\alpha)}}{\xi(1-\rho^2)} \xexpr \yexpr \\ &
- \frac{e^{-y}}{2(1-\rho^2)} \xexpr^2
\end{split}
\label{Disc_Lag}
\end{equation}
which can be simplified to
\begin{equation}
\begin{split}
\hat{L} = &-\frac{e^{-y}}{2(1-\rho^2)} \left( \xexprs - \frac{\rho
e^{y(3/2-\alpha)}}{\xi} \left( \yexprs \right) \right)^2\\ & -
\frac{e^{2y(1-\alpha)}}{2\xi^2} \yexpr^2
\end{split}
\end{equation}
This Lagrangian is difficult to deal with analytically and hence
we will consider ways to obtain numerical algorithms for the
problem.

It should be emphasized that the above Lagrangian is strictly only
correct in the limit $N \tendsto \infty$ and includes terms of
order $O(\epsilon)$ and greater apart from the above expression.

\section{Discretized Version of the Action} The action is defined
as \begin{equation} S = \int Ldt.
\end{equation} The discretized version of the action is given
by $S = \epsilon \sum_{i=1}^N L_i + O(\epsilon)$ where $L_i$ is
the Lagrangian at time step $i$. The propagator can be written in
terms of the action as
\begin{align}
\matel{x, y}{e^{-\hat{H}\tau}}{x'} &= \int_{-\infty}^\infty dy'
\matel{x, y}{e^{-\hat{H}\tau}}{x', y'}\\ &= \lim_{N \tendsto
\infty} \int DX DY e^S \label{dispro}
\end{align}
where we define
\begin{align*}
DX &= \frac{e^{-y_N/2}}{\sqrt{2\pi\epsilon (1-\rho^2)}}
\prod_{i=1}^{N-1} \int_{-\infty}^\infty \frac{dx_i
e^{-y_i/2}}{\sqrt{2\pi\epsilon (1-\rho^2)}}\\ DY &=
\int_{-\infty}^\infty dy_0 \left(\prod_{i=1}^{N-1}
\int_{-\infty}^\infty \frac{dy_i e^{y_i(1-\alpha) }}{\xi
\sqrt{2\pi\epsilon} } \right)
\end{align*}
(again $x_0 = x',\, x_N = x,\, y_0 = y'$ and $y_N = y$). We note
that the action is quadratic in $x$. This enables us to integrate
over the stock price.

We define
\begin{equation}
Q = \int DX e^{S_x} = \frac{e^{-y_N/2}}{\sqrt{2\pi\epsilon
(1-\rho^2)}} \prod_{i=1}^{N-1} \int_{-\infty}^\infty \frac{dx_i
e^{-y_i/2}}{\sqrt{2\pi\epsilon (1-\rho^2)}} e^{S_x}
\end{equation}
which is the integral of the action over the stock price.

We now find $Q$. The $x$\/-dependent term in the Lagrangian is
\begin{equation}
L_x(i) = -\frac{e^{-y_i}}{2(1-\rho^2)} \left( \xexprsi -
\frac{\rho e^{y_i(3/2-\alpha)}}{\xi}\yexpri \right)^2
\end{equation}
Let
\begin{equation}
c_i = r - \frac{e^{y_i}}{2} - \frac{\rho e^{y_i(3/2-\alpha)}}{\xi}
\yexpri
\end{equation}
Hence,
\begin{equation}
S_x = -\frac{1}{2\epsilon(1-\rho^2)} \sum_{i=1}^N e^{-y_i} (x_i -
x_{i-1} + \epsilon c_i)^2
\end{equation}
We now change the variables to $z_i$ defined by
\begin{equation}
x_i = z_i - \epsilon \sum_{j=1}^i c_i
\end{equation}
Then,
\begin{equation}
S_z = -\frac{1}{2\epsilon(1-\rho^2)} \sum_{i=1}^N e^{-y_i}
(z_i-z_{i-1})^2
\end{equation}
obtaining
\begin{equation}
Q = \int DX e^{S_x} = \frac{e^{-y_N/2}}{\sqrt{2\pi\epsilon
(1-\rho^2)}} \prod_{i=1}^{N-1} \int_{-\infty}^\infty \frac{dz_i
e^{-y_i/2}}{\sqrt{2\pi\epsilon (1-\rho^2)}} e^{S_z}
\end{equation}
All the $z_i$ integrations can be performed exactly by a process
of induction. The exact procedure can be found in any textbook on
path integration such as Kleinert\cite{kleinert} or
Roepstorff\cite{roep}. (It is also treated in Baaquie\cite{baaq}).
We illustrate the method below.

The integration over $z_1$ is easily performed. We obtain
\begin{equation}
\begin{split}
&\int_{-\infty}^\infty \frac{dz_1e^{-{y_1/2}}}{\sqrt{2\pi\epsilon
(1-\rho^2)}} \exp\left( -\frac{1}{2\epsilon (1-\rho^2)}
[e^{-y_2}(z_2-z_1)^2 + e^{-y_1}(z_1-z_0)^2]\right)\\ &=
\frac{e^{y_2/2}}{\sqrt{e^{y_1}+e^{y_2}}} \exp\left(
-\frac{1}{2\epsilon (1-\rho^2)}\frac{1}{e^{y_1}+e^{y_2}}
(z_2-z_0)^2 \right)
\end{split}
\end{equation}

The above integration can be repeatedly performed over all the
variables $z_i$ to obtain
\begin{equation}
Q = \frac{e^{S_1}}{\sqrt{2\pi\epsilon (1-\rho^2) \sum_{i=1}^N
e^{y_i}}}
\end{equation}
where
\begin{equation}
\begin{split}
S_1 &= -\frac{1}{2\epsilon (1-\rho^2) \sum_{i=1}^N e^{y_i}}
(z_N-z_0)^2\\ &= -\frac{1}{2\epsilon (1-\rho^2) \sum_{i=1}^N
e^{y_i}} (x-x'+\epsilon\sum_{i=1}^N c_i)^2\\ &=
-\frac{1}{2\epsilon (1-\rho^2) \sum_{i=1}^N e^{y_i}} \times \\
&\left[ x-x'+\epsilon\sum_{i=1}^N \left\{
r-\frac{e^{y_i}}{2}-\frac{\rho e^{y_i(3/2-\alpha)}}{\xi}
\yexpri\right\} \right]^2. \label{sone}
\end{split}
\end{equation}
On taking the limit, $N \tendsto \infty$, we get
\begin{equation}
S_1 = -\frac{1}{2(1-\rho^2) \omega} \left( x-x' + r\tau -
\frac{\omega}{2} - \frac{\rho (e^{y(\tau)(3/2-\alpha)} -
e^{y(0)(3/2-\alpha)})}{(3/2-\alpha) \xi}- \frac{\rho
\lambda}{\xi}\theta - \frac{\rho \mu}{\xi}\eta + \frac{\rho
\xi}{2} \zeta \right)^2 \label{Beautiful}
\end{equation}
The term $e^{y(0)(3/2-\alpha)}$ arises from the fact that
$\int_0^\tau dt e^{y(3/2-\alpha)} \frac{dy}{dt} =
\int_{y(0)}^{y(\tau)} dy e^{y(3/2-\alpha)}$ and
\begin{align}
\omega = &\int_0^\tau e^y dt = \int_0^\tau Vdt\\ \theta =
&\int_0^\tau e^{y(1/2-\alpha)} dt = \int_0^\tau V^{1/2-\alpha}
dt\\ \eta = &\int_0^\tau e^{y(3/2-\alpha)} dt = \int_0^\tau
V^{3/2-\alpha} dt\\ \zeta = &\int_0^\tau e^{y(\alpha-1/2)} dt =
\int_0^\tau V^{\alpha-1/2} dt.
\end{align}
Hence, if we can find the joint probability density functions for
$\omega,\, \theta,\, \eta,\, \zeta$ and $\nu = V^{3/2-\alpha}(0)$,
with $V$ following the stochastic process in (\ref{proc2}), we
obtain an analytic solution for the problem given by
\begin{equation}
\matel{x, y}{e^{-\hat{H}\tau}}{x'} = \int_0^\infty d\omega
\int_0^\infty d\theta \int_0^\infty d\eta \int_0^\infty d\zeta
\int_0^\infty d\nu \frac{e^{S_1(\omega,\, \theta,\, \eta,\, \zeta,
\nu)}}{\sqrt{2\pi\epsilon (1-\rho^2) \omega}} f(\omega,\,
\theta,\, \eta,\, \zeta,\, \nu) \label{Beautifulone}
\end{equation}
where $f$ is the joint probability distribution function.

Hence, we retain the discrete solution which finally gives us
\begin{eqnarray}
\matel{x, y}{e^{-\hat{H}\tau}}{x'} & = & \int DY
\frac{e^{S_0+S_1}}{ \sqrt{2\pi\epsilon (1-\rho^2) \sum_{i=1}^N
e^{y_i}}} \label{propagator} \\ & \equiv & \int DY P(x, y, \tau |
x^\prime, y^\prime)
\end{eqnarray}
where $S_1$ is given in (\ref{sone}) and
\begin{align}
S_0 &= -\frac{\epsilon}{2\xi^2} \sum_{i=1}^N
e^{2y_i(1-\alpha)}\yexpri^2\\ DY &= dy_0 \left( \prod_{i=1}^{N-1}
\frac{dy_i e^{y_i(1-\alpha)}}{\sqrt{2\pi \epsilon}\xi}\right)
\label{szero}
\end{align}
We need to start from the path integral given in eq(\ref{dispro})
to study numerically correlators such as $< \expon^{X_n}
\expon^{Y_m}> $. However, if one is interested solely in the price
of the option, one needs to determine only $P(x, y, \tau |
x^\prime, y^\prime)$ and in this case, the simplified
discrete-time path integral obtained in eq(\ref{propagator})
should be used as the starting point for numerical studies.

%%%%%%%%%%%%%%%%%%%%%%%%%%%%%%%%%%%%%%%%%%%%%%%%%%%%%%%%%%%%%%%%


\begin{thebibliography}{99}
\bibitem{baaq} B. Baaquie, J. de Phys. I, {\bf 7} (1997) L733.
\bibitem{ball} C. Ball and A. Roma, J. Fin. Quant. Analy.,{\bf 29}
(1994) 589.
\bibitem{bhanot} G. Bhanot, C. Lang and C. Rebbi, Comp. Phys. Comm.,
(1982) 275
\bibitem{black} F. Black and M. Scholes, J. Pol. Econ., {\bf 8} (1973),
637.
\bibitem{bodurtha} J. Bodurtha and G. Courtadon, Working Paper, Ohio Sate
University, WPS 84-69 (1984)
\bibitem{bour} J. Bouchaud and G. Lori and D. Sornette, Risk Mag.
(1995).
\bibitem{cameron} R. Cameron and W. Martin, J. Math. Phys., {\bf 34}
(1944) 195.
\bibitem{cont1} R. Cont, M. Potters and J. Bouchaud, Lanl Preprint
cond-mat-9705087.
\bibitem{cox} J. Cox and S. Ross, J. Fin. Econ., {\bf 3} (1976) 145.
\bibitem{cox85} J. Cox and M. Rubenstein, {\it Options Markets} (Prentice Hall, New York,
1985).
\bibitem{sanjiv} S. Das and R. Sudaram, J. Fin. Quant. Anal. (1999), to
appear in June issue.
\bibitem{fin} T. Finucane, {\it Fourth Symposium on the Frontiers of
Massively Parallel Computers}, Mclean, Virginia.
\bibitem{heston} S. Heston, The Rev, Fin. Stud., {\bf 6} (1993) 327.
\bibitem{hull3} J. Hull and A. White, Adv. Fut. Opt. Res., {\bf 3}
(1988), 29.
\bibitem{hullone} J. Hull and A. White, J. Int. Money Fin., {\bf 6}
(1987) 131.
\bibitem{hull2} J. Hull and A. White, J. Fin., {\bf XLII}
(1987) 281.
\bibitem{jarrow} R. Jarrow and A. Rudd, {\it Option Pricing} (Irwin, New York,
1983).
\bibitem{johnson} H. Johnson and D. Shanno, J. Fin Quant. Anal., {\bf 22}
(1993) 143.
\bibitem{jones} E. Jones, J. Fin. Econ.,{\bf 13} (1984) 91
\bibitem{kleinert} H. Kleinert, {\it Path Integrals in Quantum Mechanics,
Statistics and Polymer Physics} (World Sci., Singapore, 1990).
\bibitem{kon} S. Kon, J. Fin., {\bf 39} (1984) 147.
\bibitem{lamoureaux} C. Lamoureaux and W. Lastrapes, The Rev Fin. Stud.,
{\bf 6} (1993) 293.
\bibitem{merton1} R. Merton, Bell J. Econ. Management Sci., {\bf 4}
(1973) 141.
\bibitem{merton1a} R. Merton, J. Fin. Econ., {\bf 3} (1976) 125.
\bibitem{mill} K. Mills, M. Vinson and G. Cheng, Syracuse University
NPAC Technical Report SCCS-260, 1993.
\bibitem{pott1} M. Potters, R. Cont and J. Bouchaud, Lanl Preprint
cond-mat-9609172.
\bibitem{roep} G. Roepstorff, {\it Path Integral Approach to Quantum
Physics} (Springer Verlag, Berlin, 1991).
\bibitem{rubensteinone} M. Rubenstein, J. Fin., {\bf 38} (1983) 213.
\bibitem{rubenstein} M. Rubenstein, J. Fin., {\bf 40} (1986) 445.
\bibitem{scott} L. Scott, J. Fin Quant.Anal., {\bf 22} (1987) 419.
\end{thebibliography}
\end{document}